\documentclass[useAMS,usenatbib]{mn2e}
\usepackage{journals}
\usepackage{times}
\usepackage{float}

\newcommand{\sub}[1]{_{\mathrm{#1}}}
\newcommand{\sur}[1]{^{\mathrm{#1}}}
\newcommand{\erg}[1]{$10^{#1}$~erg}
\newcommand{\ergs}[1]{$10^{#1}$~erg~s$^{-1}$}
\newcommand{\ergscm}[1]{$10^{#1}$~erg~cm$^{-2}$~s$^{-1}$}
\newcommand{\ergcm}[1]{$10^{#1}$~erg~cm$^{-2}$}

\def\rin{$r_{\rm m}$}

\def\rc{$r_{\rm c}$}

\usepackage{url}
\usepackage{amssymb}
\usepackage{graphicx}
\usepackage{textcomp}

\title[The RB Type II burst population]{A population study of type II bursts in the Rapid Burster}
\author[Bagnoli et al.]{T. Bagnoli$^{1,2}$\thanks{E-mail: t.bagnoli@sron.nl},
J.J.M. in 't Zand$^{1}$,
C.R. D'Angelo$^{3}$
and D.K. Galloway$^{4}$\thanks{Also at School of Physics and Astronomy, Monash University, Clayton, VIC 3800, Australia.}
\\
$^{1}$SRON Netherlands Institute for Space Research,
Sorbonnelaan 2, 3584 CA Utrecht, The Netherlands\\
$^{2}$Astronomical Institute ``Anton Pannekoek'', University of Amsterdam,
Postbus 94249, 1090 GE Amsterdam, The Netherlands\\
$^{3}$Leiden Observatory, Leiden University, Postbus 9513, NL-2300 RA Leiden, the Netherlands\\
$^{4}$Monash Centre for Astrophysics (MoCA), Monash University,
Clayton, VIC 3800, Australia
\\
\\
\\
\textup{Accepted 2015 February 12.  Received 2015 February 10; in original form 2014 December 19}
}

\begin{document}

\date{}

\pagerange{\pageref{firstpage}--\pageref{lastpage}} \pubyear{}

\maketitle

\label{firstpage}

\begin{abstract}
Type II bursts are thought to arise from instabilities
in the accretion flow onto a neutron star in an X-ray binary.
Despite having been known for almost 40 years, no model
can yet satisfactorily account for all their properties.
To shed light on the nature of this phenomenon
and provide a reference for future theoretical work,
we study the entire sample of
\textit{Rossi X-ray Timing Explorer} data
of type II bursts from the Rapid Burster (MXB 1730--335).
We find that type II bursts are Eddington-limited in flux,
that a larger amount of energy goes in the bursts
than in the persistent emission,
that type II bursts can be as short as 0.130~s,
and that the distribution of recurrence times
drops abruptly below 15--18~s.
We highlight the complicated feedback
between type II bursts and the NS surface thermonuclear explosions
known as type I bursts, and between type II bursts and the
persistent emission.
We review a number of models for type II bursts.
While no model can reproduce all the observed burst properties
and explain the source uniqueness,
models involving a gating role for the magnetic field
come closest to matching the properties of our sample.
The uniqueness of the source may be explained by a special combination
of magnetic field strength, stellar spin period
and alignment between the magnetic field and the spin axis.
\end{abstract}

\begin{keywords}
stars: neutron -- X-rays: binaries -- X-rays: bursts -- X-rays: individual: MXB 1730-335
\end{keywords}

\section{Introduction}

X-ray bursts observed in
accreting neutron stars (NS) in low-mass X-ray binaries (LMXBs)
come in two types, sometimes similar in appearance
($\sim 1$ min time scales, \ergs{38} peak luminosities)
but thought to be due to very different mechanisms,
because of the large difference (by factors $\sim 10^2$) in
average power.

Type I bursts arise from the heating and cooling of the NS photosphere
after a thermonuclear shell flash of accreted material (see reviews by
\citealt*{1993SSRv...62..223L}, \citealt{2006csxs.book..113S}, \citealt{2008ApJS..179..360G}).
This is consistent with thermal spectra
of varying temperature and roughly constant
emitting area.
Despite some remaining open questions
(e.g., \citealt{2003A&A...405.1033C}, \citealt*{2011A&A...527A.139S}, \citealt{2012ARA&A..50..609W})
a general understanding of the physics underlying type I bursts has been reached,
also thanks to the wide range of circumstances
in which they occur (type of LMXB, accretion rates, etc)
provided in particular by the relatively large amount of bursters (over a hundred).

Type II bursts remain largely a puzzle.
Soon after their discovery,
\citet*{1978Natur.271..630H} showed that their
short recurrence times rule out the possibility
that they are powered by nuclear burning,
ascribing the source instead to gravitational potential energy.
Little more is understood about their origin, also because
they are only confidently detected in two sources:
the Rapid Burster (MXB 1730--335, hereafter RB),
a recurring transient which is the object of the present study
and which also exhibits type I bursts,
and the Bursting Pulsar (BP), a slowly rotating (2~Hz)
accretion-powered pulsar \citep{1996Natur.379..799K}.

One of the first properties of type II bursts
discovered in the RB is the relaxation-oscillator behaviour:
the fluence in a type II burst
is proportional to the waiting time to the next burst \citep{1976ApJ...207L..95L}.
This is not observed in the BP \citep{1996Natur.379..799K}.
The relaxation-oscillator behaviour
seems to point to the presence of a mass storage
from which matter can be released and then replenished at a constant rate,
so that the amount of depleted material
determines the time to refill the buffer \citep{1993SSRv...62..223L}.
The accretion disc is the most obvious candidate
for this buffer and all proposed models involve one
(e.g., \citealt{1992ApJ...385..651W,1993ApJ...402..593S}).
The various time scales at different positions in the disc
can help to explain the range of observed type II burst time scales,
while a key role in acting as a gating mechanism might be played by a magnetic field
\citep[e.g., recent simulations by][]{2008MNRAS.386..673K,2010MNRAS.406.1208D,2012MNRAS.421...63R,2012MNRAS.420..416D}.

Notwithstanding that models
involving some role for the NS magnetic field seem promising,
the strength of the magnetic field is unfortunately not known in the RB
due to the lack of pulsations.
For the BP, estimates of the surface strength
of the dipolar component of the magnetic field $B$ put it 
in the range $(2-6)\times 10^{10}$~G \citep{2014ApJ...796L...9D}.
While this is larger than in typical NS LMXBs
\citep{1993SSRv...62..223L},
it is not unique among bursters
\citep[estimates put the magnetic field of the NS in IGR J17480-2446
at $10^9-10^{10}$~G; e.g.,][]{2011ApJ...740L...8C}.
Therefore, at least another parameter
besides the magnetic field strength
(e.g., the stellar spin, or the inclination
of $B$ with respect to the stellar spin axis)
must determine the type II burst occurrence.

\citet[][hereafter B13]{2013MNRAS.431.1947B}
have shown that thermonuclear (type I) X-ray bursts in the RB
occur over an unusually large range of accretion luminosities,
from about 45 per cent of the Eddington luminosity $L\sub{Edd}$
(assuming it is located in the globular cluster Liller 1
at a distance of $7.9 \pm 0.9$~kpc, \citealt*{2010MNRAS.402.1729V})
down to almost quiescence \citep{2000A&A...363..188M}.
However, while the type I bursts
span the entire range of luminosities,
type II bursts only appear below
a critical luminosity of about 10 per cent $L\sub{Edd}$ \citep[][B13]{1999MNRAS.307..179G}.
Hence, there are a type-I-only phase and a mixed phase.

The type-I-only state is a soft state because
the persistent emission spectrum is dominated by a disc black-body component,
and it spans the range 10--45 per cent of $L\sub{Edd}$.
This is almost unique among type I burst sources:
all other bursters show nearly no type I bursts
above $L \sim 0.1 L\sub{Edd}$ \citep{2003A&A...405.1033C},
except for the slow (11~Hz) pulsar IGR J17480--2446 \citep{2012ApJ...748...82L}.
The abundance of type I bursts in the RB above this threshold
prompted B13 to propose that it is a slow rotator too,
and to speculate that this might actually be
one of the necessary ingredients for producing type II bursts.

As the RB exits the soft state,
the first type II bursts appear.
These first type II bursts are the most energetic, long and infrequent.
A brief (typically a few days) intermediate state exists in which
these are accompanied by double-peaked type I bursts and a QPO
at 0.25~Hz \citep[hereafter B14]{2014MNRAS.437.2790B}.
The unusual shape of the type I bursts
found in this short-lived intermediate outburst phase
seems to be due to a temporary obscuration of the NS surface
during the burst decay, producing the double peaks.
The ensuing type II bursts, the obscuration phenomenon and the occurrence of a QPO
are probably related to the change in the accretion geometry
that is widely thought to take place
between the soft and the hard state
\citep*[e.g., review by][and references therein]{2007A&ARv..15....1D}.

Having studied the soft and intermediate states in our aforementioned papers,
we now complete our study of the entire
\textit{Rossi X-ray Timing Explorer} (\textit{RXTE}, 1995--2012) RB database
by focussing on the type-II-burst dominated hard state,
so called because of the power-law component
that dominates the persistent emission spectrum.
We aim at constructing a population study
of type II bursts
to compare it with the predictions of the few available theoretical models,
and as a future reference.
In Sec.~\ref{sec:obs} we introduce the dataset,
and in Sec.~\ref{sec:ana} we illustrate our methodology.
In Sec.~\ref{sec:res} we present the type II burst properties and
their relationship with the persistent emission
and the type I bursts,
taking advantage of the unprecedented detail provided by the \textit{RXTE} data.
We summarize our findings, some of which were previously unknown,
in Sec.~\ref{sec:sum}, and
discuss their physical implications in Sec.~\ref{sec:dis}.
Finally, in Sec.~\ref{sec:mod} we review proposed theoretical models,
checking their predictions against the properties of our sample.

	\section{Observations}\label{sec:obs}

\begin{figure*}
  	\includegraphics[width=2.0\columnwidth]{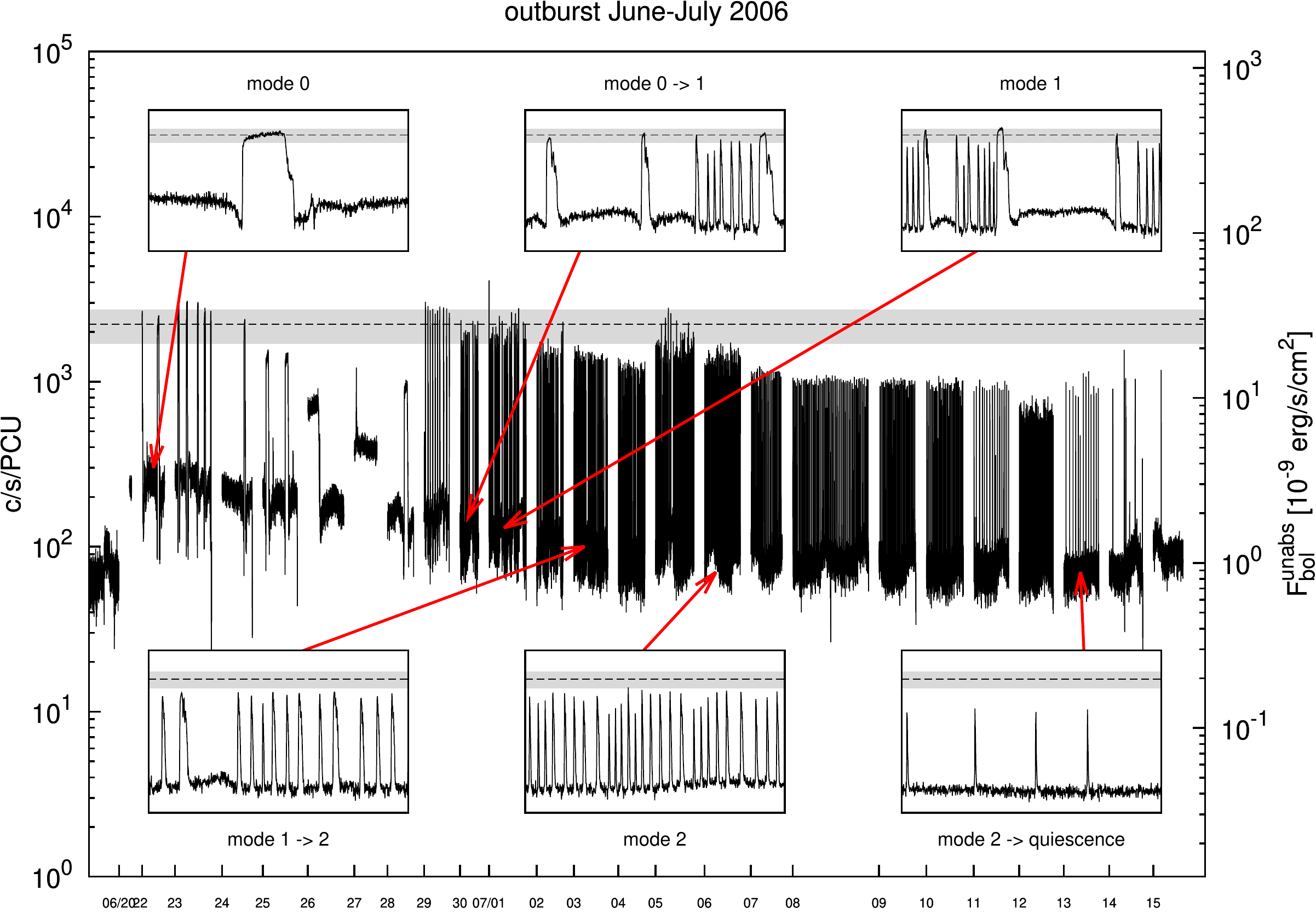}\\
  		\caption{\small{
		Sandwiched light curve of the observations
		of the outburst starting on June 20 2006,
		for the complete PCA bandpass at 1~s resolution resolution.
		Each consecutive data stretch is typically about 3000~s long.
		Data gaps mark jumps in time when there were no observations.
		Photon count rate per PCU is on the left y-axis,
		unabsorbed bolometric flux on the right y-axis.
		The horizontal dashed line indicates the Eddington luminosity,
		with the grey-shaded area indicating the uncertainties on the distance and
		the spectral fit (see Sec.~\ref{sec:proxy}).
		Count rates are not background subtracted in this plot,
		but are corrected for the number of active PCUs
		and the collimator response.
		(throughout the rest of the paper,
		the background counts have been subtracted).
		4U 1728-34 was never in the field of view.
		The non-zero count rates before the onset of the outburst
		are therefore due to the instrumental background
		and the galactic ridge emission
		(see Fig.~2 in \citealt{1998ApJ...505..134V}).
		The insets show the varying
		(following in time left to right, top to bottom) type II bursting modes
		during the outburst decay. Each inset shows 1~ks of data, plotted logarithmically
		between 50 and 5000~c~s$^{-1}$PCU$^{-1}$.
		This outburst shows the prototypical sequence
		mode 0 $\rightarrow$ mode 1 $\rightarrow$ mode 2 in bursting behaviour.
		The soft state, however, does not appear
		at the beginning of the outburst,
		hence preceding the type II bursting phase as in most cases,
		but it is briefly entered by the RB around June 27,
		giving rise to the long flat peak around that date.
		Note that the shorter plateau visible on June 26 is actually
		the longest type II burst in our sample, at about 1100~s
		(see Sec.~\ref{sec:typeIIbursts}).
		}}
	\label{fig:outburst92026}
\end{figure*}

\subsection{The general sample and burst identification}\label{sec:sample}

This study is about measurements with
the Proportional Counter Array (PCA) onboard \textit{RXTE}.
It consists of five co-aligned proportional counter units (PCUs)
that combine to a total photon collecting area of 
8000~cm$^2$ over a 2 to 60~keV bandpass \citep{2006ApJS..163..401J}.

Nearly all \textit{RXTE} observations of the RB
(exposure time 2.4~Ms)
were carried out in one of three configurations.
For 553~ks, the PCA was directly pointed at the source.
These observations are contaminated by the presence of
the persistently bright 4U 1728-34
at an angular distance of 0.56$\degr$.
For 612~ks, the pointing was offset by just this amount,
thus avoiding the contamination at the expense of roughly half the effective area.
Finally, for 1.3~Ms the PCA was pointed at 4U 1728-34,
serendipitously encompassing the RB.
We have investigated all of the 2.4~Ms of data.

We extracted a lightcurve from the '\textsc{Standard-1}' data for all the observations
containing the RB in the field of view (FOV).
These data have 0.125~s time resolution and
no photon energy resolution.
The telescope orientation was determined
with the \textsc{ftool pcaclrsp},
and deemed stable if the jitter was below 20~arcsec.

We performed a visual inspection of all the light curves.
For the details of source (RB vs 4U 1728-34)
and burst (type I vs type II) identification,
we refer the reader to Appendix~\ref{sec:appendix_search}.
In total, we identified 123 type I
and 8458 type II bursts from the RB.

\subsection{Type II bursting modes}\label{sec:outburst}

As already mentioned, type II bursts are only emitted by the RB at persistent luminosities
below 10 per cent of the Eddington limit
\citep[B13,][]{1999MNRAS.307..179G}, when the RB is in the hard state (B14).
Actually the RB almost never reaches higher luminosities during outbursts after 1999
(\citealt{2002A&A...381L..45M}; B14).

Three modes of type II bursting behaviour
follow one another in a smooth transition,
differing in lightcurves, energetics and bursting patterns
\citep{1979ApJ...227..555M,1999MNRAS.307..179G}.
To illustrate the evolution of the type II burst behaviour,
we plot the light curve from \textit{RXTE} data 
of the outburst in June-July 2006 (Fig.~\ref{fig:outburst92026}).

The first type II bursts are hundreds of seconds long, often with flat peaks,
and have the longest recurrence times
(sometimes longer than a few ks observation).
The persistent emission drops right before and after them.
These are the so-called mode-0 type II bursts (Fig.~\ref{fig:outburst92026}, top-left inset).
We identified a record long mode 0 burst with a duration
of at least 1100~s.
It is visible in Fig.~\ref{fig:outburst92026}
as the long plateau on June 26.

At lower persistent fluxes,
the large bursts are increasingly preceded by
sequences of 8 to 40 short bursts.
After the final longer burst, a longer gap is measured,
during which the persistent emission rises to a level
higher than that of the intra-burst emission,
in a ``hump'' shape.
This bursting pattern is called mode 1 (Fig.~\ref{fig:outburst92026}, top-right inset).

As the outburst decays further, the sequences of mode 1 type II bursts become increasingly long,
the final bursts less energetic, and the humps shorter and less pronounced.
Eventually all bursts have similar energetics and durations: this is the so-called mode 2
(Fig.~\ref{fig:outburst92026}, bottom-middle inset).
Finally, even these bursts become increasingly sparse,
as the source moves to quiescence (Fig.~\ref{fig:outburst92026}, bottom-right inset).

We caution that the sequence ``no type II bursts $\rightarrow$ mode 0
$\rightarrow$ mode 1 $\rightarrow$ mode 2''
is only a \textit{general} trend.
For instance, the flat peak in Fig.~\ref{fig:outburst92026}
starting on June 27 is not a very long mode 0 type II burst,
but an actual switch of the source to the soft state,
during which a type I burst can be seen,
but no type II bursts.
Two observations were taken the day before and the day after
showing long mode 0 type II bursts,
meaning that the soft state lasted briefly.

\begin{figure}
  	\includegraphics[width=\columnwidth]{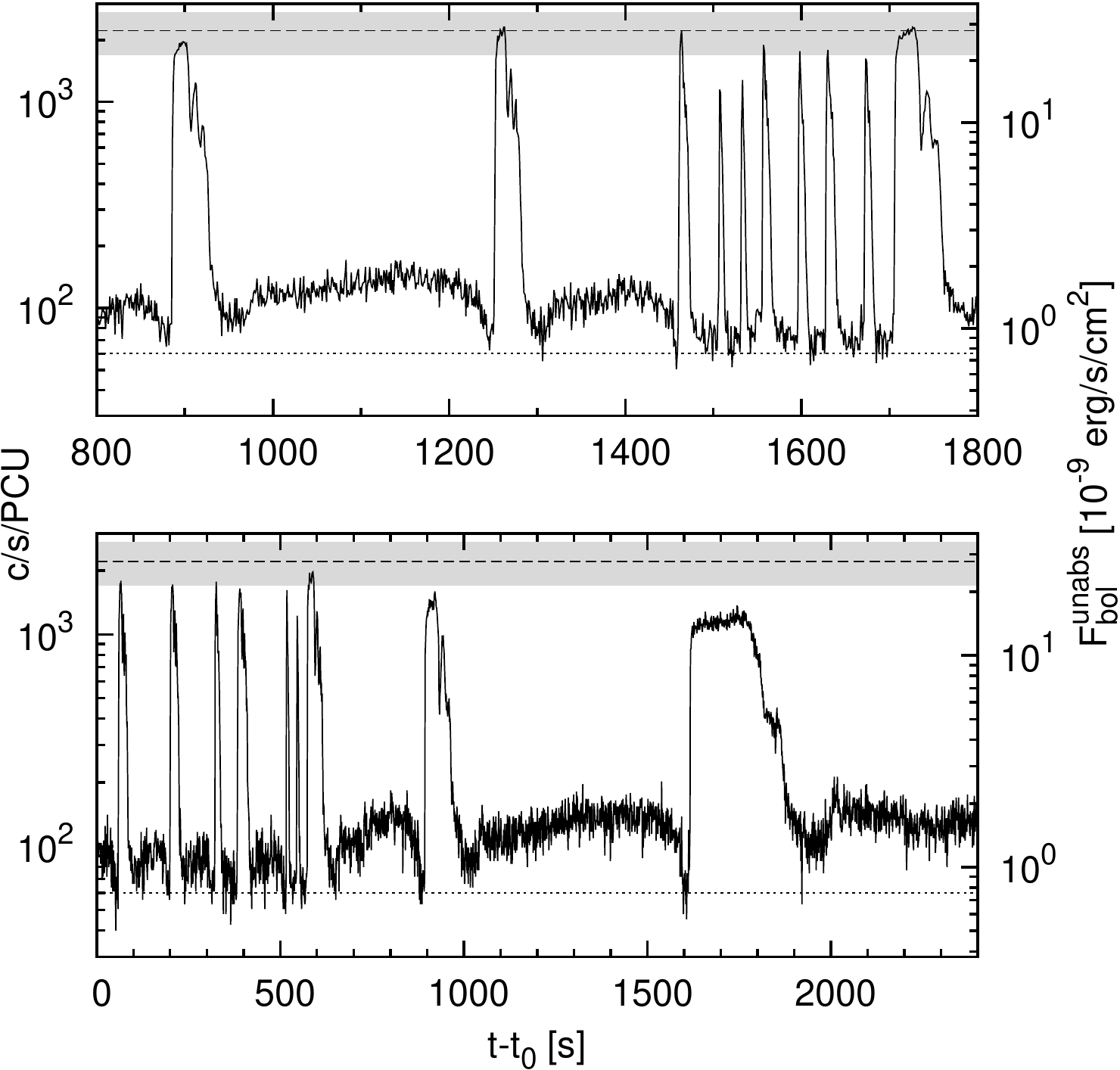}
  		\caption{\small{Examples of mode switching,
		0 $\rightarrow$ 1 (top panel) and 1 $\rightarrow$ 0 (bottom panel).
		Time axis with respect to the start of the ObsID.
		The y-axes are as in Fig.~\ref{fig:outburst92026}.
		The horizontal dashed line indicates the Eddington luminosity,
		with the grey-shaded area indicating the uncertainties on the distance and
		the spectral fit (see Sec.~\ref{sec:proxy}).
		The dotted line shows $F\sub{pers}\sur{min} =$ 60~c~s$^{-1}$PCU$^{-1}$.
		The quiescent level between quick bursts is the same as the dip level
		before and after mode 0 bursts.
		The RB therefore does not ``dip'' but rather shows humps
		of enhanced emission after long type II bursts.
		}}
	\label{fig:mode_switching}
\end{figure}

Also, the RB can sometimes be seen switching back and forth
between modes.
The switch is sudden rather than gradual
\citep[as shown in Fig.~\ref{fig:mode_switching}, already noted by][]{1993SSRv...62..223L},
although the RB can be seen to transit back and forth
between modes for a few ks.
It is clear from the picture that the minima before and after a mode 0 burst
are not ``dips'':
rather, the humps in between are what is peculiar and
are a feature common to all very energetic type II bursts, independent of bursting mode.

The sampling along an outburst is often patchy,
meaning that it is not easy to establish an average duration for each mode.
The outburst in Fig.~\ref{fig:outburst92026}
is probably the best sampled one.
It features the longest mode 0 phase in the entire dataset, lasting $\approx 8.7$ days,
albeit including a brief return into the soft state.
The longest uninterrupted mode 0 bursting phase
occurred in an earlier outburst, and lasted $\approx 6.5$ days.
Mode 1 bursts are visible in Fig.~\ref{fig:outburst92026}
for $\approx 2.1$ days.
In a later outburst, they were observed to last for twice that time,
$\approx 5.2$ days.
Finally, mode 2 burst were continuously observed for up to 24 days,
the longest stretch in the dataset.

\begin{figure*}
  	\includegraphics[width=0.66\columnwidth]{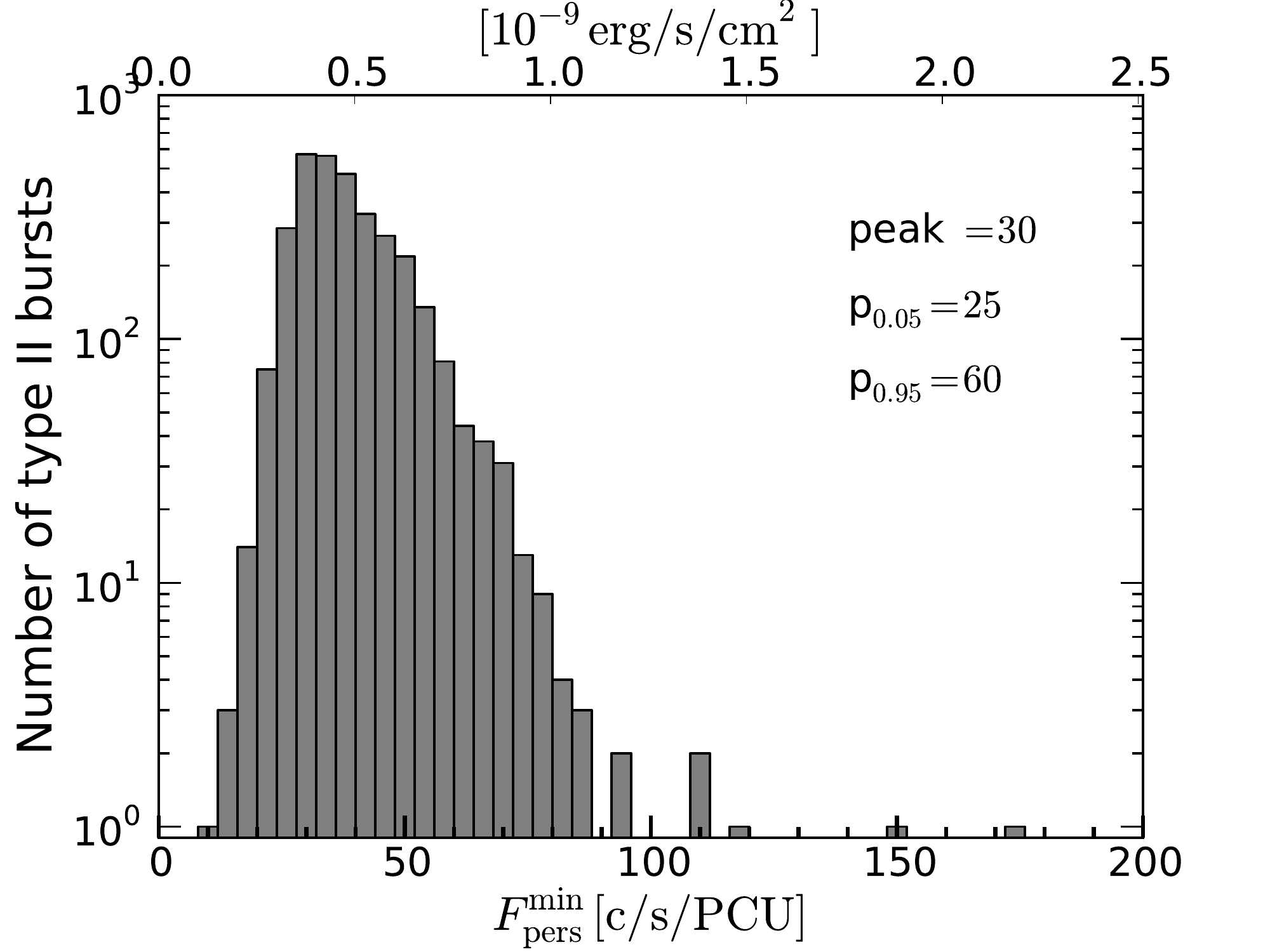}
  	\includegraphics[width=0.66\columnwidth]{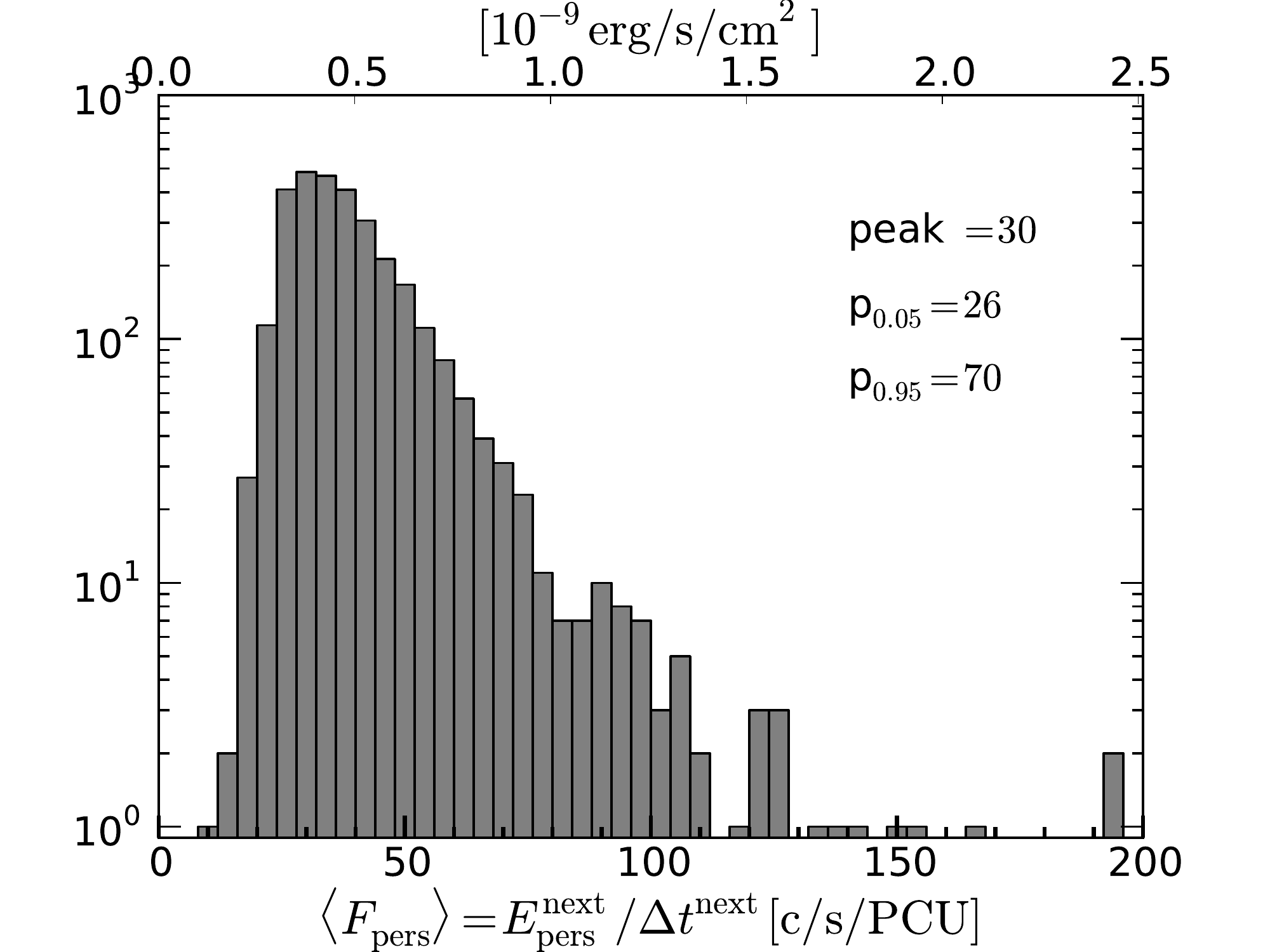}
  	\includegraphics[width=0.66\columnwidth]{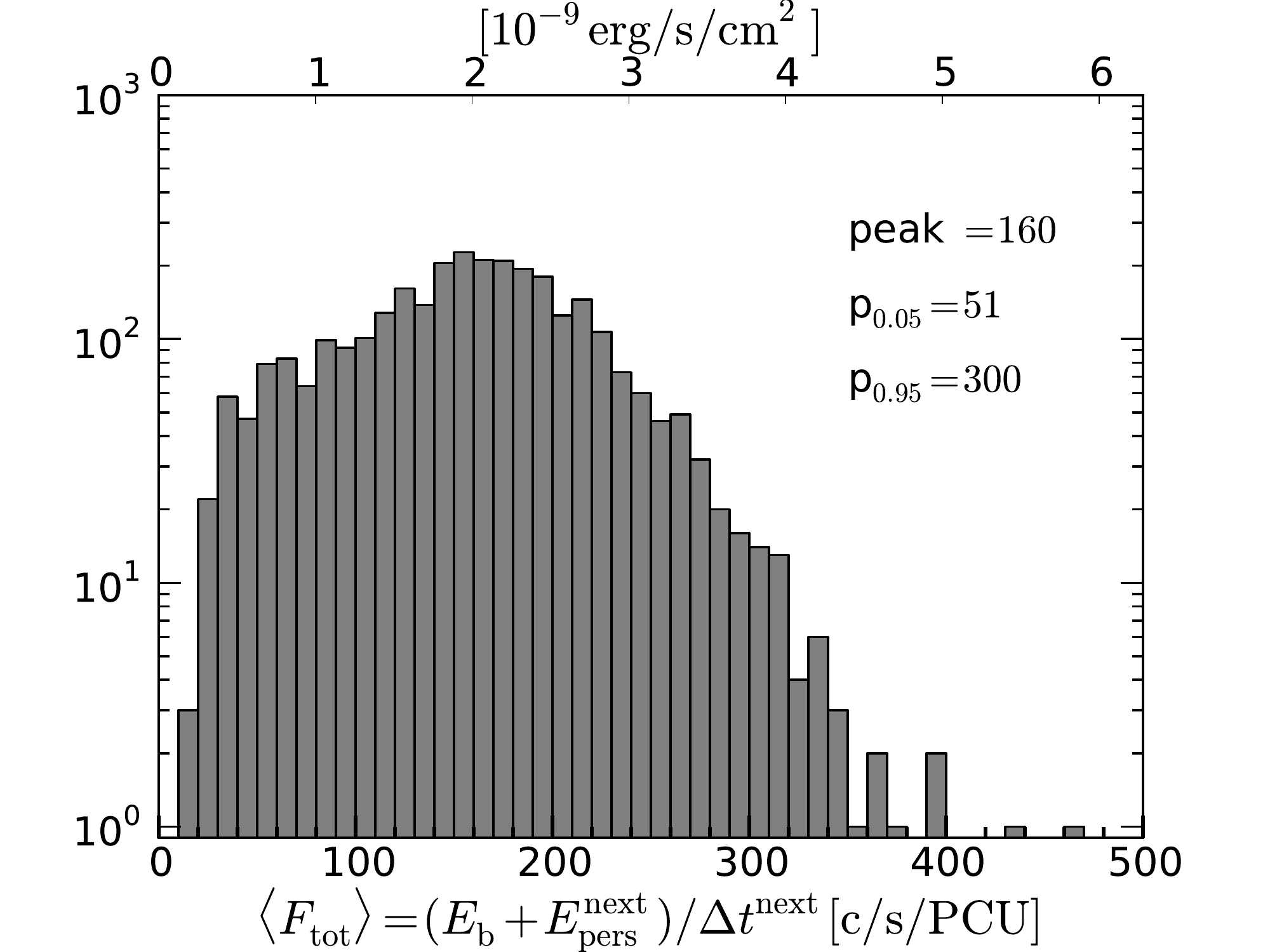}
  		\caption{\small{Histograms of all persistent flux parameters (see Sec.~\ref{sec:persistent}):
			the persistent count rate immediately before a burst $F\sub{pers}\sur{min}$,
			the average persistent count rate after a burst $\langle F\sub{pers}\rangle$
			and the total average flux (including the burst) $\langle F\sub{tot}\rangle$.
			The peak of the distribution, 5th and 95th percentiles are indicated for each parameter.
			Only bursts that were not contaminated by 4U~1728-34 have been included (see Sec.~\ref{sec:sample}).}}
	\label{fig:persistent_hist}
\end{figure*}

\subsection{The type II burst sample}

Our search routine found 8458 type II bursts in the entire dataset
(see Sec.~\ref{sec:sample}).
The results are summarized in Table~\ref{table:bursts}.
We determined the instrumental 
and cosmic diffuse background with the \textsc{ftool pcabackest}.
No bursts were analyzed that are outside so-called good time intervals
(GTI), are affected by data gaps
are overlapping with type I bursts (from either the RB or 4U 1728-34),
or are in ObsIDs for which the collimator response was not constant
because of a slew.
This left us with 7601 bursts.\footnote{A table including all the burst properties
for the entire sample is available online.}
Of these, 3662 type II bursts are present in 389~ks of direct pointings,
while 3183 were observed during 341~ks of offset pointings,
and 756 during 92~ks of 4U 1728-34 observations.

\begin{table}
 \caption{The sample of analyzed$\sur{(a)}$ type II bursts.}\label{table:bursts}
 \begin{tabular}{@{}lcccc}
   \hline\hline
Bursting mode$\sur{(b)}$	&	\multicolumn{3}{c}{Number of bursts per aimpoint$\sur{(c)}$}	&Total\\
	&RB	&	offset	& 1728-34 &\\
   \hline
0	& 78	& 24	& 2	& 104\\
1	& 964	& 895	& 663	& 2522\\
2	& 2620	& 2260	& 91	& 4971\\
unclear & 0	& 4	& 0	& 4\\
   \hline
all	& 3662	& 3183	& 756	& 7601\\	
    \hline\hline
 \end{tabular}

 \medskip
$\sur{(a)}$About 800 bursts identified by our routine were not analyzed,
see Sec.~\ref{sec:sample}.\\
$\sur{(b)}$For a description of the bursting patterns, see Sec.\ref{sec:outburst}.\\
$\sur{(c)}$The different orientations of the telescope in the data
and their effect on FOV contamination and effective area are discussed in Sec.~\ref{sec:sample}.
\end{table}

In total, the RB was stably in mode 0 during the entirety of an ObsID for 114~ks,
in mode 1 for 170~ks, and in mode 2 for 492~ks.
In 51~ks of data,
the behaviour was intermediate between two consecutive modes,
switching often back and forth during the transition.
We found 104 mode-0 bursts,
2522 mode-1 bursts
and 4971 mode-2 bursts.
For 4 bursts taking place during very short observations (a few hundred seconds)
we could not clearly assign a mode.

Furthermore, during 210~ks of data, the RB was at such a high luminosity
that only type I bursts were observed (B13).
Finally, no RB bursts of either type were found in ObsIDs totalling 1.4~Ms
(mostly in observations with the instrument pointed at 4U 1728-34),
when the RB was in quiescence.

\section{Methodology}\label{sec:ana}

\subsection{Counts as proxy for energy}\label{sec:proxy}

We adopted the photon counts as a proxy for the energy flux,
rather than fitting time-resolved spectra.
This approach offers an obvious advantage
in speed of the data reduction.
Furthermore, the intra-burst persistent emission
would often be too short and weak to constrain the spectrum well.
Finally, the flux variations,
which can be very large during type II bursts,
are largely achromatic;
it is mostly the normalization factor
and not the intrinsic spectral shape that is changing.
This is shown in Appendix~\ref{sec:spectralstudy},
where we perform time-resolved spectroscopy
on a representative selection of the type II bursts.
We obtained the best fits with a Comptonization model
\citep[\textsc{Comptt} in {\sc XSpec}][]{1994ApJ...434..570T}.
The conversion factor between PCA counts and 3--25~keV fluence is
$1.02\times$\ergcm{-11} per c/PCU.
The uncertainty induced by the spectral shape is 10 per cent (full range, not $1 \sigma$).
The bolometric correction would be $1.24 \pm 0.07$ for the \textsc{Comptt} model.

For a ``canonical'' NS (mass $M = 1.4 M_{\odot}$)
the Eddington luminosity $L\sub{Edd}$ is $2.1\times$\ergs{38}.
This assumes isotropic emission, no gravitational redshift
and an accreted H fraction $X = 0.7$
because the RB type I bursts show long (100~s) tails
that are typical of burning of H-rich material.
For a distance of $(7.9\pm0.9)$~kpc, this yields a flux $F\sub{Edd} = (2.8 \pm 0.3)\times$\ergscm{-8}.
Given the above conversion to the bolometric flux,
and propagating the relative errors on the spectral shape,
bolometric correction and distance,
the Eddington luminosity in the RB
corresponds to $F\sub{Edd}\sur{PCA} = (2200\pm 500)$~c~s$^{-1}$PCU$^{-1}$.

\subsection{Extracting the sample properties}\label{sec:variables}

For each type II burst,
we calculated from the \textsc{Standard-1} light curve
the start time and duration $t\sub{dur}$ at 0.125~s accuracy
by seeking backward and forward in time for the first
2 consecutive time bins
where the count rate was below the persistent emission level.
This way, we also determined $\Delta t\sur{prev}$ and $\Delta t\sur{next}$,
the recurrence times with respect to the previous and the next burst respectively,
as the difference between the start times.
Because the burst duration can be a significant fraction
of the duty cycle for type II bursts (unlike for most type I bursts),
the actual non-bursting time
can be significantly shorter than the recurrence times so defined.

We defined two different measurements of the intra-burst persistent flux.
First, we calculated the actual intra-burst fluence $E\sub{pers}$
and divided that by the intra-burst time,
providing the average persistent flux $\langle F\sub{pers}\rangle$.
In a vast majority of cases the instantaneous flux
was stable during the interval, and deviated little from the average.

As mentioned in Sec.~\ref{sec:outburst},
the persistent emission however tends to rise after very energetic bursts
(the mode 0 and end-of-sequence mode 1 bursts),
reaching a minimum immediately before and after those bursts.
Therefore, for the purpose of measuring the net burst emission properties,
we also calculated the average count rate $F\sub{pers}\sur{min}$ 
in the 10~s before a burst (or less, when $\Delta t\sur{prev}$ is shorter).
This assumes that, during the burst,
the persistent emission will be stable at the minimum level reached
right before it.

We made this choice for two reasons.
Firstly, this assumption has long been part of the standard approach
to burst analysis \citep[e.g.,][]{2008ApJS..179..360G}.
Secondly, the reappearance of the persistent emission
at roughly the same flux after the energetic bursts
(see e.g., Fig.~\ref{fig:mode_switching})
supports our choice.

We subtracted $F\sub{pers}\sur{min}$
from the observed count rates
to calculate the net burst emission properties.
We defined the peak flux $F\sub{peak}$ and the total burst fluence $E\sub{b}$
as the maximum net count rate and the integrated burst net counts during this interval.

The ratio of $E\sub{pers}$ to $E\sub{b}$ is the parameter $\alpha$,
which is key to distinguishing between type I and II bursts,
because the ratio of the fluence liberated in the persistent emission
to that the burst diagnoses whether the latter can be explained by thermonuclear burning
of the accreted material or not.
For reasons that will become clearer in Sec.~\ref{sec:typeIIbursts},
we defined two values of $\alpha$ for each burst,
employing $E\sub{pers}$ \textit{before} and
\textit{after} the burst: thus we retrieved
a backward-defined $\alpha\sur{prev}$
and a forward-defined $\alpha\sur{next}$.
The previous is the parameter that is normally employed in studies of type I bursts,
but as we will see, the latter shows a more significant relationship for type II bursts.

We subtracted the instrumental background counts
(calculated and interpolated from \textsc{Standard-2} data in 16~s resolution)
from all quantities, and then normalized them
by the number of active PCUs and the correction
given by the telescope orientation (see Sec.~\ref{sec:sample}).
Clearly, all quantities related to the persistent emission have to be considered
upper limits (lower for $\alpha$) when the FOV is contaminated by 4U~1728-34.
We did not include these in our analysis.

	\section{Results}\label{sec:res}

\subsection{The persistent emission}\label{sec:persistent}

\begin{figure}
  	\includegraphics[width=\columnwidth]{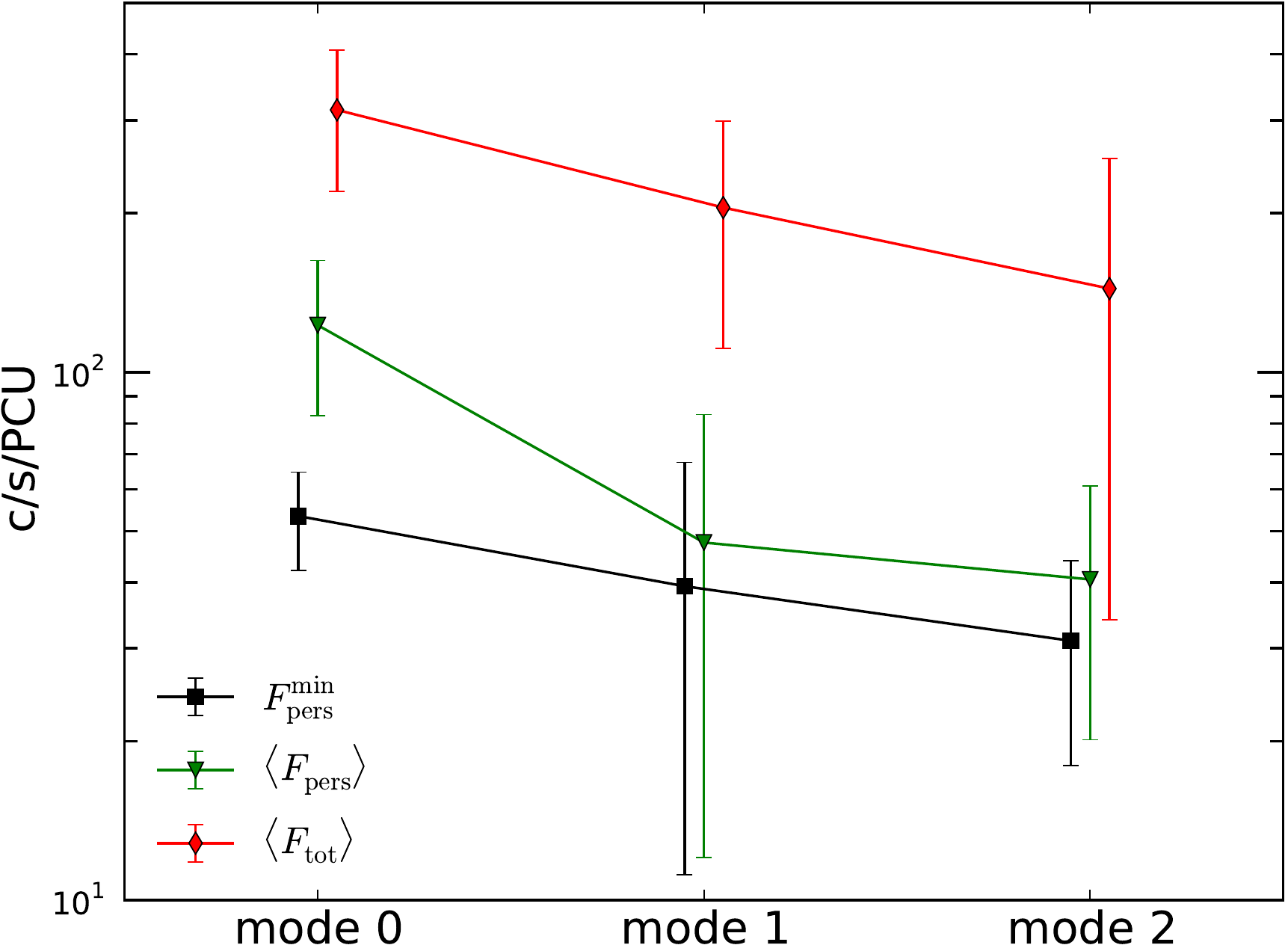}
  		\caption{\small{The average values of $F\sub{pers}\sur{min}$,
		$\langle F\sub{pers}\rangle$ and $\langle F\sub{tot}\rangle$
		during mode 0, 1 and 2 type bursts (see Sec.~\ref{sec:persistent}).
		The error bar indicates the standard deviation in each sample.}}
	\label{fig:persistent_trend}
\end{figure}

\begin{figure*}
  	\includegraphics[width=\columnwidth]{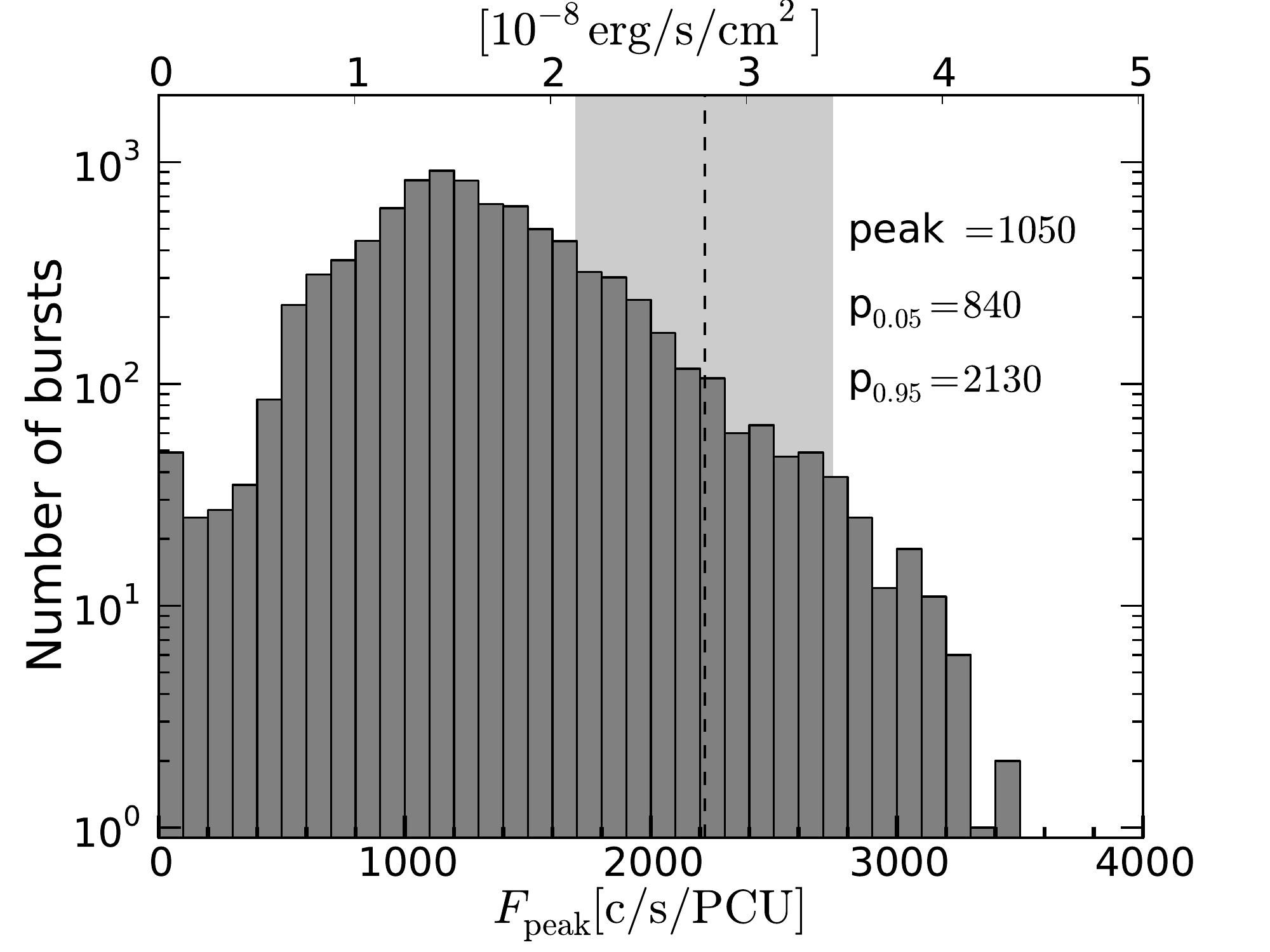}
  	\includegraphics[width=\columnwidth]{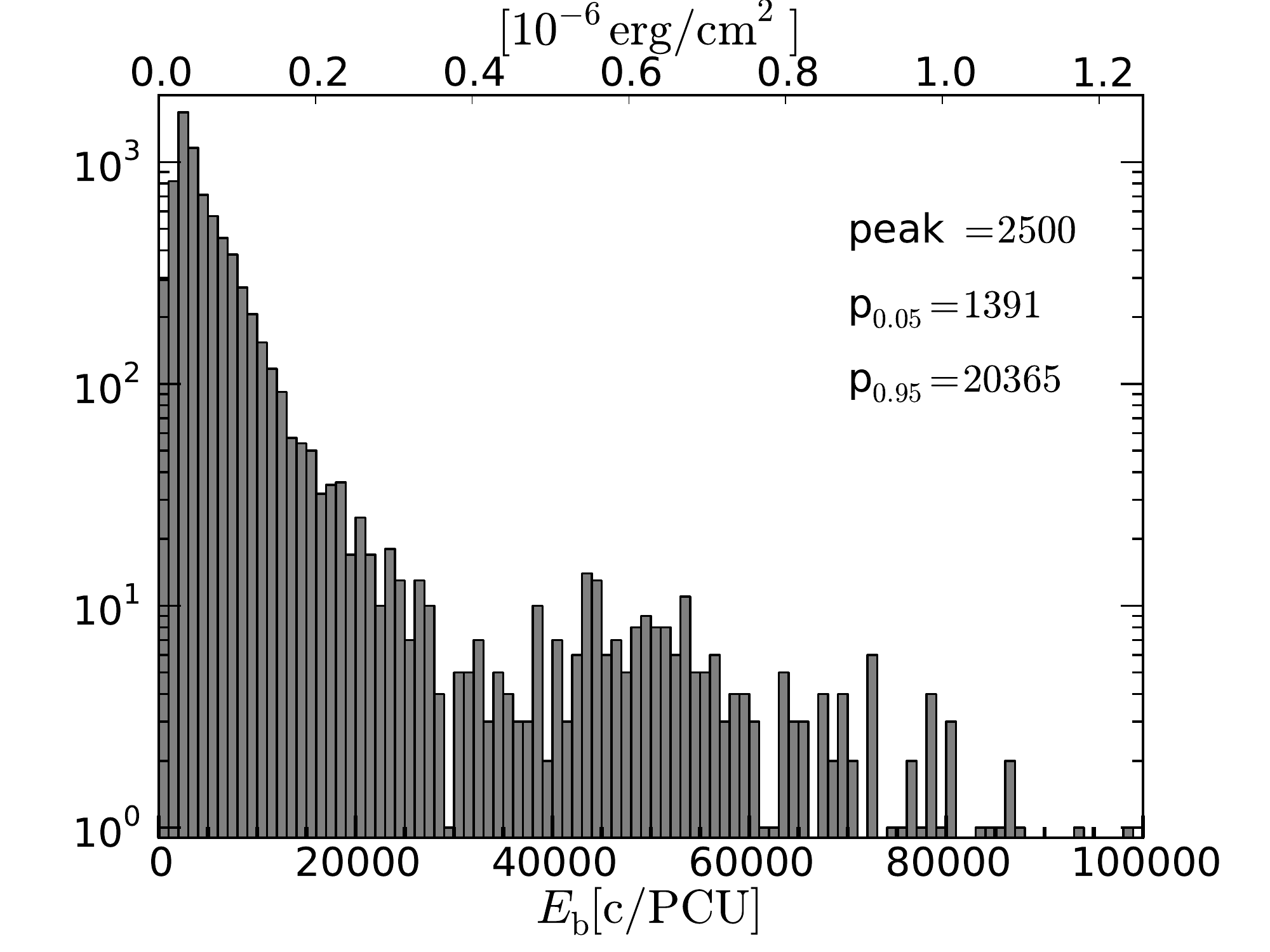}\\
  	\includegraphics[width=\columnwidth]{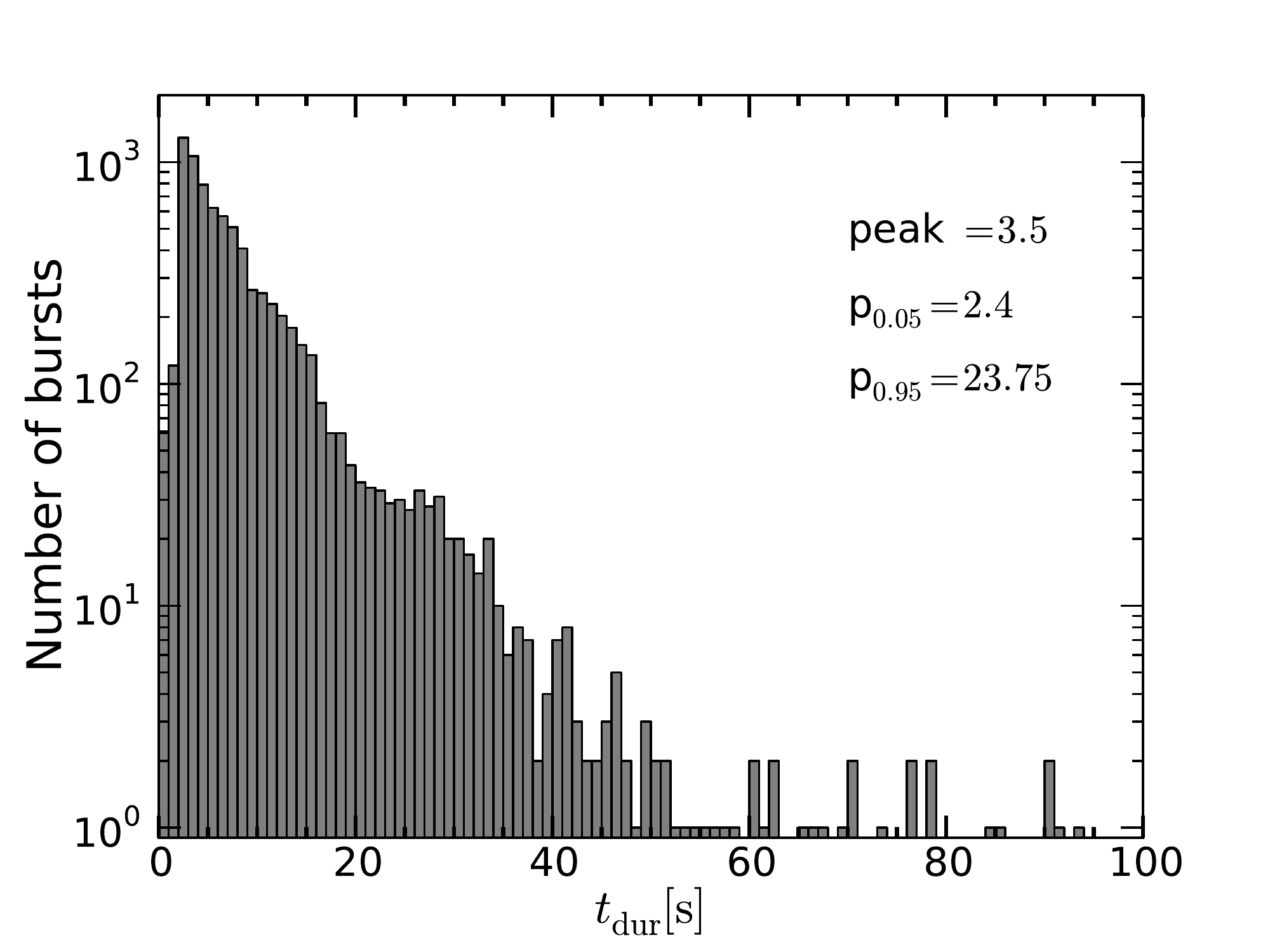}
  	\includegraphics[width=\columnwidth]{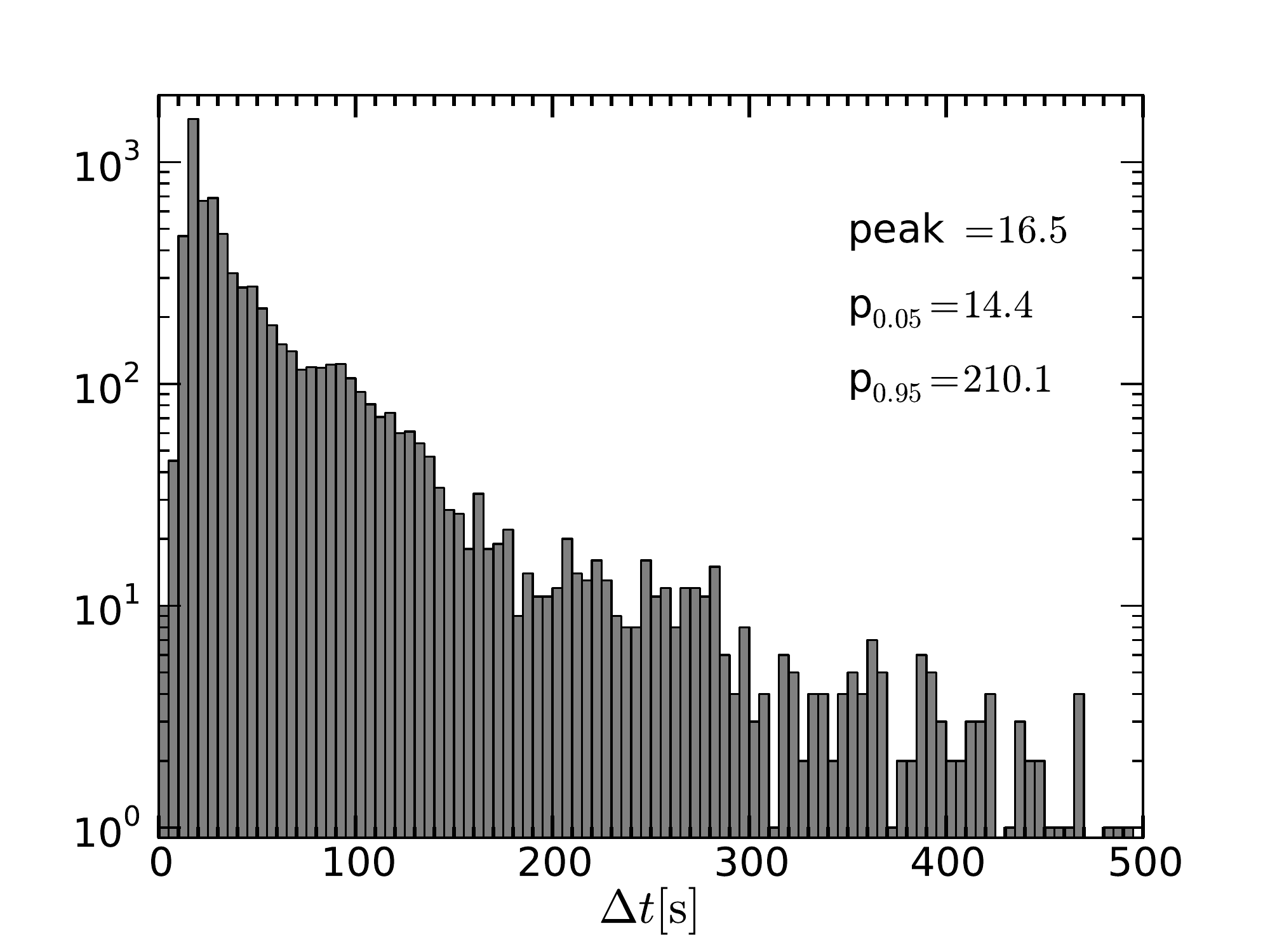}\\
  	\includegraphics[width=\columnwidth]{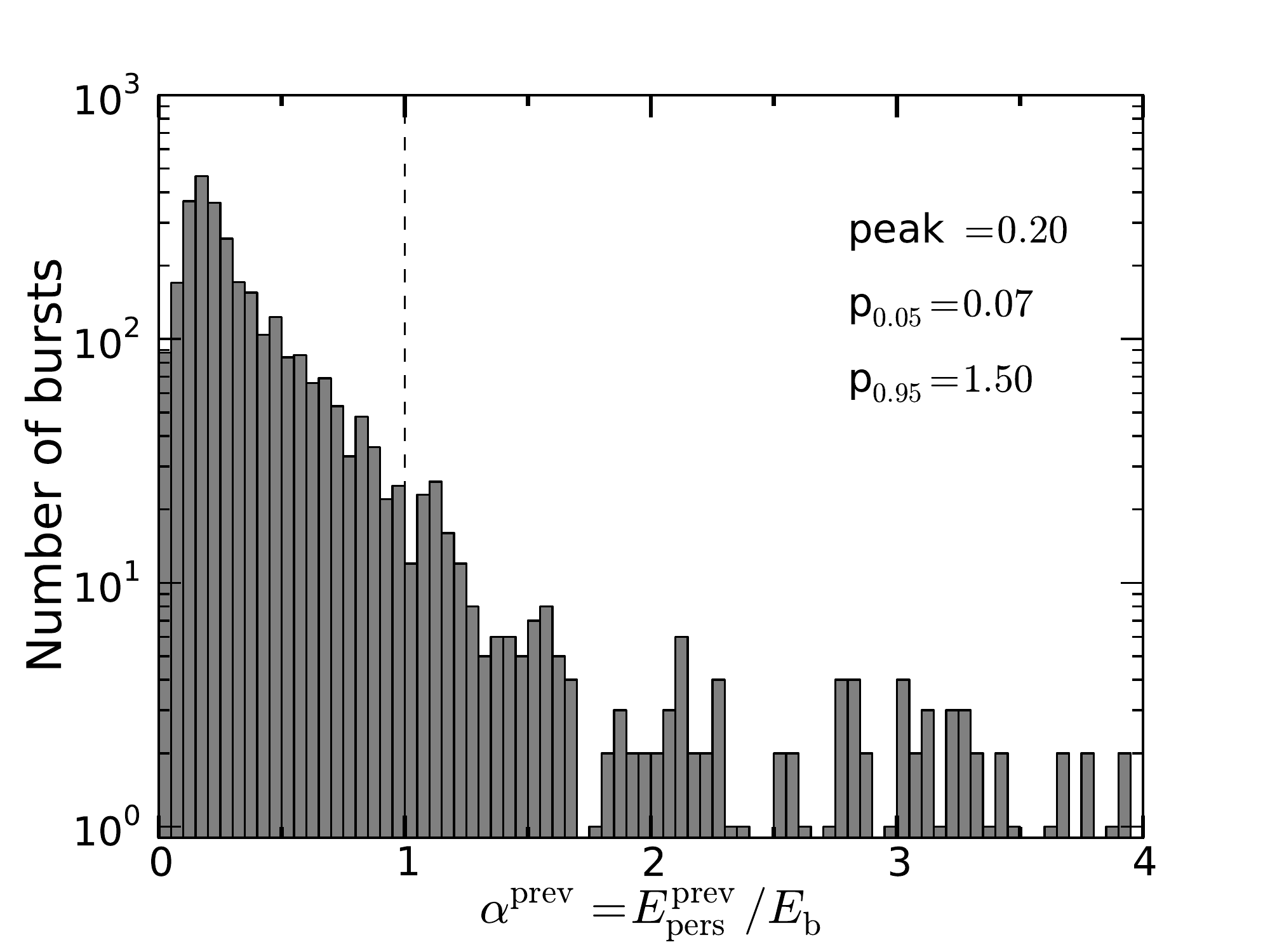}
  	\includegraphics[width=\columnwidth]{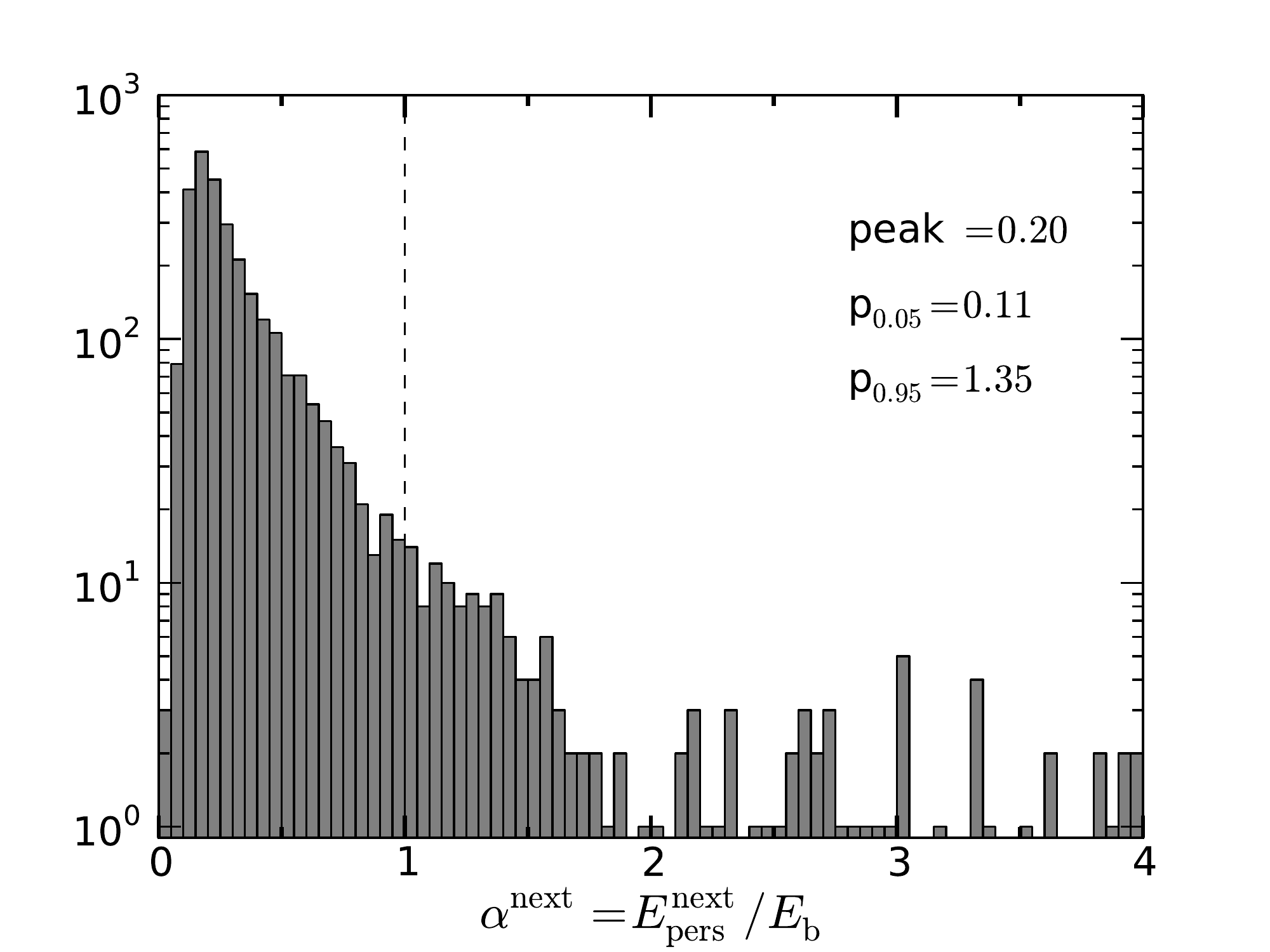}
  		\caption{\small{Histograms of all burst parameters.
		The peak of the distribution,
		5th and 95th percentiles are indicated for each parameter.
		The dashed line in the left-top plot is the flux corresponding to the
		Eddington luminosity for $M = 1.4 M_{\odot}$, $X = 0.73$.
		The grey area reflects the uncertainty in the spectral fit (see Appendix~\ref{sec:spectralstudy}).
		The shortest observed $\Delta t = 2$~s, $t\sub{dur} = 0.125$~s.
		The dashed line in the $\alpha$ plots divides the regions where
		the burst fluence exceeds that of the persistent emission and viceversa.
		$\alpha$ values have only been computed for observations
		that were not contaminated by 4U 1728-34 (see Sec.~\ref{sec:sample}).}}
	\label{fig:burst_hist}
\end{figure*}

First we look at the properties of the persistent emission
when the RB is emitting type II bursts,
whenever not contaminated by 4U 1728-34 (see Sec.~\ref{sec:sample}).
We plot the various properties in Fig.~\ref{fig:persistent_hist},
with the peak of the distribution (the highest bin)
and the 5 and 95 percentiles indicated.

The average persistent emission count rate before a burst $F\sub{pers}\sur{min}$
is limited to a narrow range of values, 
with 90 per cent of the distribution
between 25 and 60~c~s$^{-1}$PCU$^{-1}$,
or $\simeq (0.01-0.025) F\sub{Edd}$.
Therefore, no bursts are observed on top of a null net persistent emission.
The largest observed value is 110~c~s$^{-1}$PCU$^{-1}$,
corresponding to $\simeq 0.05 F\sub{Edd}$
(note that at times when there are no type II bursts,
the persistent flux can be as high as $\simeq 0.45 F\sub{Edd}$).

As explained in Sec.~\ref{sec:variables},
this is only a lower limit on the actual intra-burst emission.
However, short mode-1 and mode-2 bursts make up the majority
of our sample, so that the following histogram,
of the average persistent count rates after a burst
$\langle F\sub{pers}\rangle$,
shows a very similar distribution,
with a slightly longer tail.
The minimum values are similar to those of $F\sub{pers}\sur{min}$
because the persistent emission never dips between bursts,
but only right before and after them.
The maximum of the distribution is larger with respect to $F\sub{pers}\sur{min}$,
with all but ten bursts showing
$\langle F\sub{pers}\rangle < 200 \textrm{~c~s$^{-1}$PCU$^{-1}$} \simeq 0.09 F\sub{Edd}$.

Since both the persistent emission and the type II bursts
probably originate from the release of gravitational energy
(although possibly through different channels),
the total average flux in the interval between two bursts $\Delta t$,
including both the burst and persistent emission,
is also a potential proxy for the actual mass accretion rate of the RB during the hard state
(under the assumption that the radiative efficiency
is similar for the persistent emission and the type II bursts).
We call this flux $\langle F\sub{tot}\rangle$,
and plot it in the following histogram in Fig.~\ref{fig:persistent_hist}.
The bursts seem to increase the accretion flux by a factor of $\sim$4,
up to an upper limit of 400~c~s$^{-1}$PCU$^{-1}$, or $ \simeq 0.17 F\sub{Edd}$.
Therefore, an average 75 per cent of the accretion goes through bursts.

All three measurements show a monotonically
decreasing trend from mode 0 to 1 to 2,
as plotted in Fig.~\ref{fig:persistent_trend}.
However, the dispersion within each sample is similar or larger
than the difference in values between modes.
We obtained the same result trying to reproduce
the same figure for the outburst plotted in Fig.~\ref{fig:outburst92026} alone.
It seems therefore that the persistent emission variations
only marginally trace the changing type II burst behavior.

\subsection{The type II bursts}\label{sec:typeIIbursts}

For all 7601 analyzed type II bursts we
study their energetics (peak flux $F\sub{peak}$ and fluence $E\sub{b}$)
and timescales (duration $t\sub{dur}$ and recurrence time $\Delta t$).
Furthermore, we investigate their relationships to the persistent emission properties
for the 3183 bursts observed when 4U 1728-34 is outside the FOV.
In Fig.~\ref{fig:burst_hist} we have plotted histograms of the burst properties,
again annotated with the peak of the distribution (the highest bin)
and the 5 and 95 percentiles.

\subsubsection{Burst properties}

Starting with the distribution of peak fluxes of type II bursts,
the peak is at about half the Eddington luminosity, or 1100~c~s$^{-1}$PCU$^{-1}$.
Overall, the distribution is relatively narrow,
with 90 per cent of the bursts peaking in the range
$(1.1-2.7)\times$\ergscm{8}.
The distribution seems compatible with being Eddington limited.
Only 1 per cent of the bursts are above the upper limit
on $L\sub{Edd}$ given by the uncertainty over the distance and the spectral fit
(see Sec.~\ref{sec:proxy};
this limit is actually more clear in Fig.~\ref{fig:burst_rels1}-\ref{fig:burst_rels3}).

A larger dynamic range of values is observed for the burst fluences,
which span about three orders of magnitude,
making for a fairly large distribution.
The majority of bursts are in mode 1 or 2,
so it is not surprising that many show a relatively
small fluence.
These bursts radiate typically between 
 $10^{-8}$ and several $10^{-7}$~erg~cm$^{-2}$.
The more energetic mode 0 bursts 
show minimum fluences of several $10^{-7}$~erg~cm$^{-2}$.
42 mode 0 bursts show $E\sub{b}>$\ergcm{-6},
peaking at $E\sub{b}\sur{max} = 5.8 \times$\ergcm{-6}.

\begin{figure}
  	\includegraphics[angle=0,width=\columnwidth]{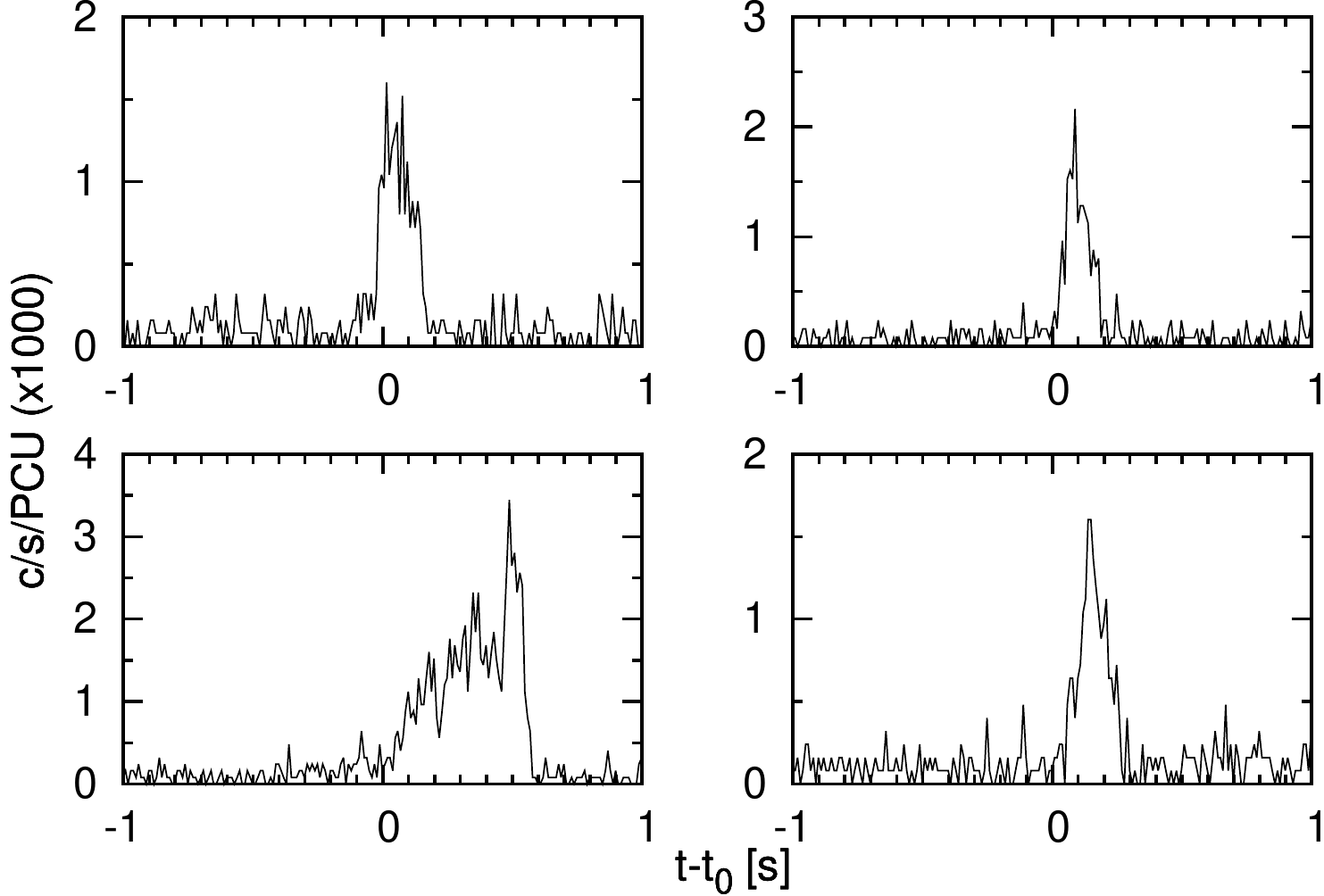}\\
\vskip 2mm
  	\includegraphics[angle=0,width=\columnwidth]{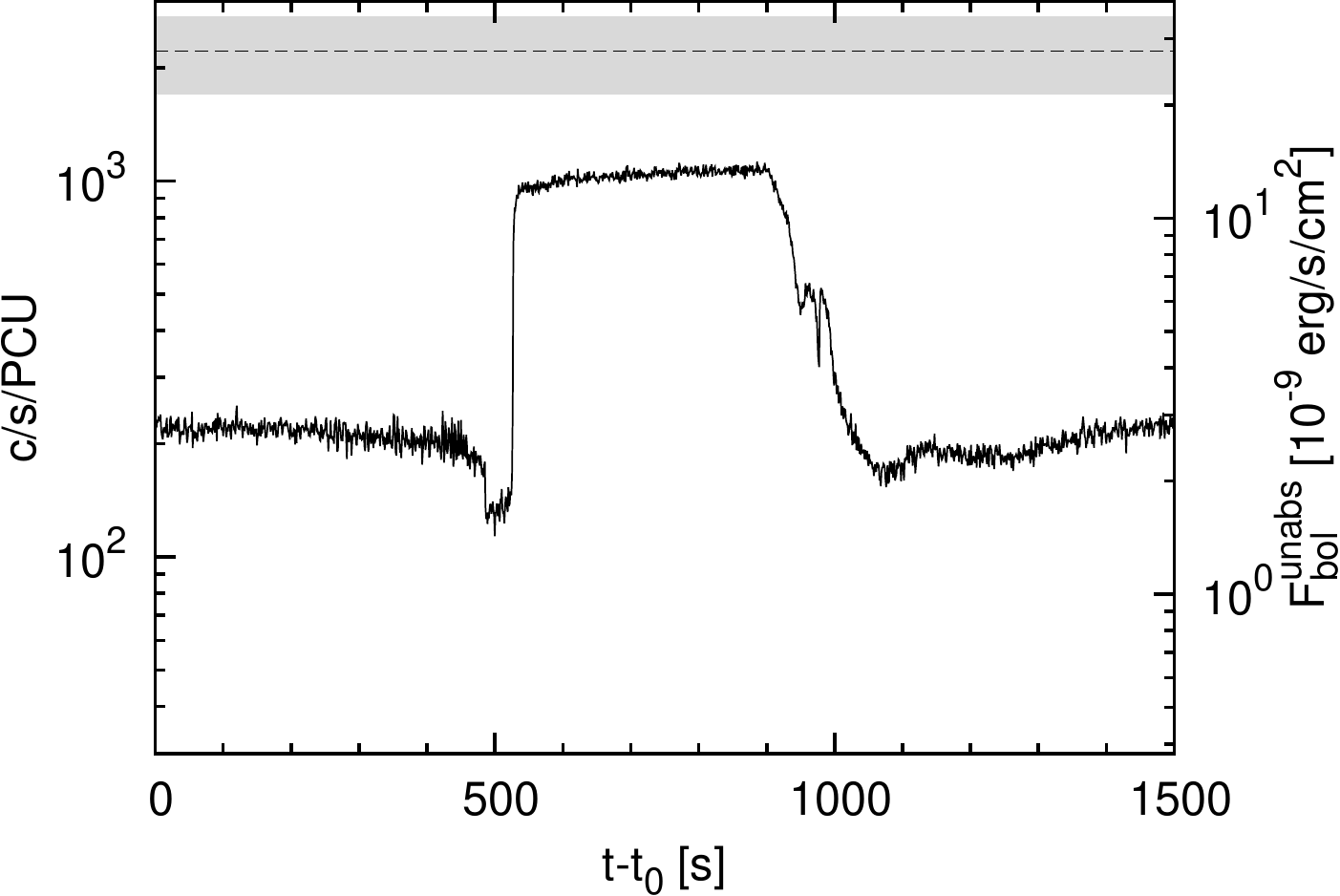}
  		\caption{\small{(\textit{Top.}) 
		Examples of some of the shortest bursts in the sample
		from the 10~ms resolution light curve of ObsID 92026-01-08-05 (July 20 2006).
		Left y-axis is as in Fig.~\ref{fig:outburst92026}.
		The bursts take place during an otherwise ordinary sequence of mode 2 type II bursts.
		(\textit{Bottom.}) For comparison, the longest
		entirely observed burst in the sample, at 533.125~s,
		from ObsID 40058-02-03-01S (April 20 1999).
		The y-axes are as in Fig.~\ref{fig:outburst92026}.
		The flat top is substantially below the Eddington limit,
		indicated by the horizontal dashed line,
		with the grey-shaded area indicating the uncertainties on the distance and
		the spectral fit (see Sec.~\ref{sec:proxy}).
		A longer burst ($\gtrsim 1100$~s, see 
		Fig.~\ref{fig:outburst92026} and Sec.~\ref{sec:typeIIbursts})
		was only partially observed.}}
	\label{fig:micro}
\end{figure}

The durations of type II bursts can vary by 4 orders of magnitude,
although 90 per cent of them last between 2.4 and 23.75~s.
The shortest bursts actually challenge the time resolution
of \textsc{Standard-1} data of 0.125~s,
and need to be resolved in event-mode data (Fig.~\ref{fig:micro}, top).
The shortest burst in our sample only lasted 130~ms.
We found 61 bursts with sub-second durations,
all during mode 2 bursting phases towards the end of an outburst,
46 of them in ObsID 92026-01-08-05 alone.

These very short bursts do not show peak fluxes
smaller than the rest of the sample.
Their fluences are however some of the smallest observed,
liberating on average $\sim 5\times$\ergcm{-9}.
Their light curves tend to show very fast rises
and equally fast decays,
although they can have shapes that are unseen
in the rest of the sample
(such as the burst in the bottom-left panel of Fig.~\ref{fig:micro},
which has a gradual rise and a sudden decay).

The longest burst observed in its entirety
lasts 533~s (Fig.~\ref{fig:micro}, bottom).
The record holder is however a burst
lasting at least 1100~s,
the rise of which was not caught in the observation.
Only 26 bursts recorded durations longer than 100~s,
all but two being mode-0 bursts.

The recurrence times are mostly below 200~s (again, with the exception of the mode-0 bursts),
the shortest and the longest (without data gaps) $\Delta t$ being 2~s and 1656~s respectively.
Overall, the distribution is extremely asymmetric.
This is logical, since the shorter the recurrence time,
the more bursts will be observed.
The peak of the distribution is however well above zero,
at $\Delta t = 15-18$~s.
Below this bin, the distribution drops abruptly.
Concerning mode 0 bursts, their recurrence times (excluding lower limits)
range between 105 and 1562~s.

For the 3183 bursts that were in observations uncontaminated by 4U 1728-34 we
compared the fluence released in the burst to that in the persistent emission.
We calculated the $\alpha\sur{prev}$ and $\alpha\sur{next}$ values, respectively,
by dividing the integrated fluence in the interval prior and after the burst
by the total burst fluence $E\sub{b}$.
The former is the parameter that is employed when dealing with type I bursts,
since thermonuclear bursts are due to the burning of material \textit{after} this has been accreted.
Fig.~\ref{fig:burst_hist} shows that the values of $\alpha\sur{prev}$ for type II bursts
peak at 0.20, meaning more energy
is released in the bursts than in the persistent emission.
They are incompatible with the values expected
from thermonuclear burning ($\simeq 40$ at least, for H rich fuel, \citealt{1993SSRv...62..223L},
and the minimum measured $\alpha \simeq 10$, \citealt{2010ApJ...718..292K}).

Even at the lowest persistent fluxes,
when the RB approaches quiescence and the type II burst frequency drops,
(see bottom-right inset of Fig.~\ref{fig:outburst92026}),
$\alpha\sur{prev} < 10$.
The only type II bursts with larger $\alpha\sur{prev}$ values
are those with sub-s durations, which can have $\alpha\sur{prev} \sim 20-200$,
but are much shorter than type I bursts.

The type II bursts are thought to draw from
the same energy reservoir as the persistent emission. 
Also, their occurrence and properties seems to affect the persistent emission that follows,
rather than to depend on the persistent emission that precedes them (see below).
Because of this, we also plot the integrated ratio of the fluence \textit{following}
a burst to the integrated burst fluence, $\alpha\sur{next}$.
The two distributions are similar, although that of $\alpha\sur{next}$ is narrower
and does not approach values as small
as the other one.
The final large bursts in mode 1 sequences
tend to have short $\Delta t\sub{prev}$ (the time interval to the previous burst),
and long $\Delta t\sub{next}$.
Hence their $\alpha\sur{prev}$ are smaller than their $\alpha\sur{next}$.

\subsubsection{Correlations}

\begin{figure}
  	\includegraphics[width=\columnwidth]{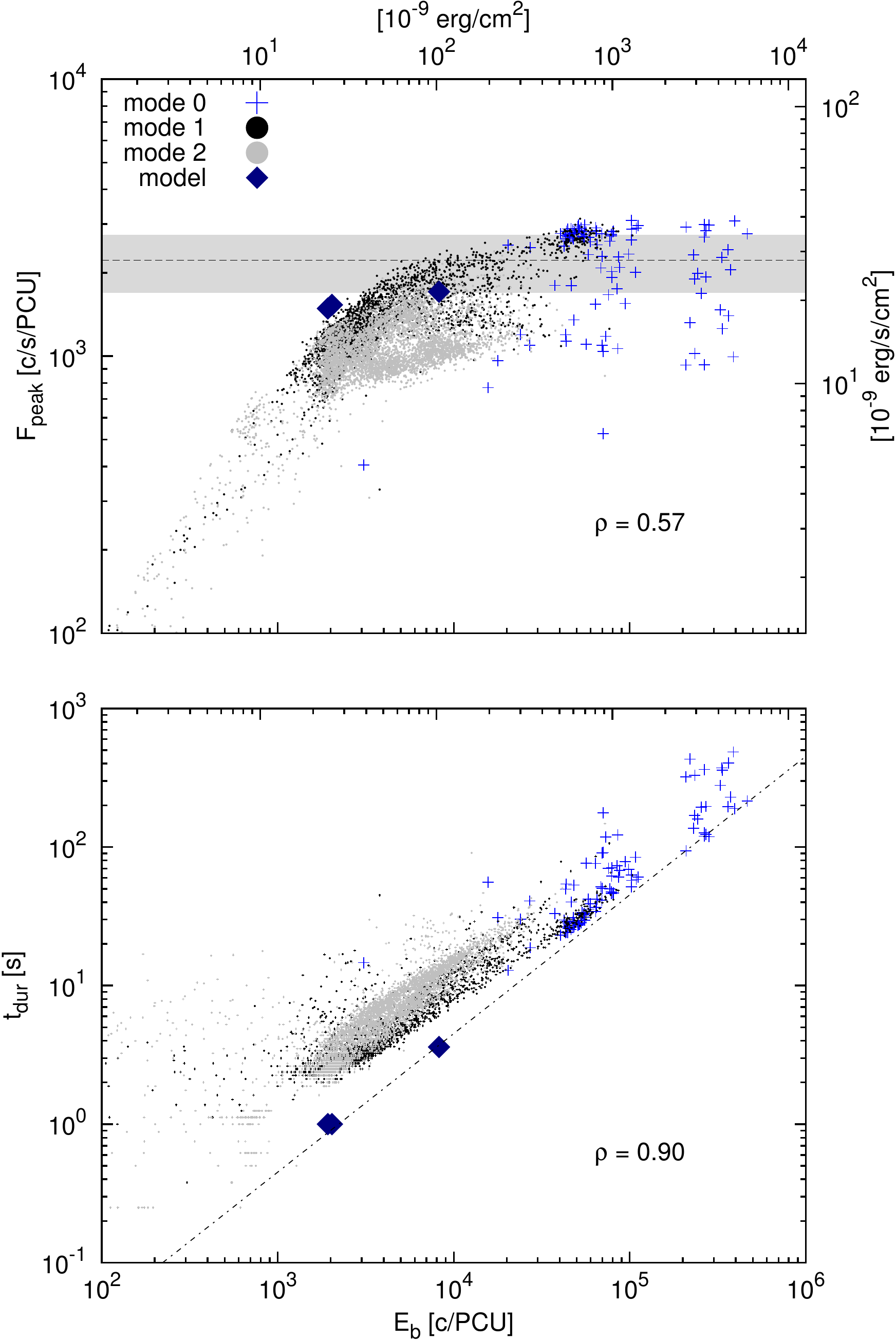}
  		\caption{\small{Peak flux $F\sub{peak}$ and burst duration $t\sub{dur}$
		as a function of the burst fluence $E\sub{b}$ (see Sec.~\ref{sec:typeIIbursts}).
		Count rates and energies are as in Fig.~\ref{fig:outburst92026}.
		The horizontal dashed line indicates the Eddington luminosity,
		with the grey-shaded area indicating the uncertainties on the distance and
		the spectral fit (see Sec.~\ref{sec:proxy}).
		The correlation (given by the Spearman correlation factor $\rho$)
		is strong in both cases,
		although it saturates in the upper plot at the flux
		corresponding to the Eddington luminosity.
		The dashed-dotted line in the bottom plot
		indicates the duration of an Eddington-limited burst
		of a given fluence.
		The red diamonds are the values predicted by three the models
		in Fig.~\ref{fig:bursts_caro}, with two of them
		nearly overlapping (see Sec.~\ref{sec:dangelo}).
		}}
	\label{fig:burst_rels1}
\end{figure}

\begin{figure}
  	\includegraphics[width=\columnwidth]{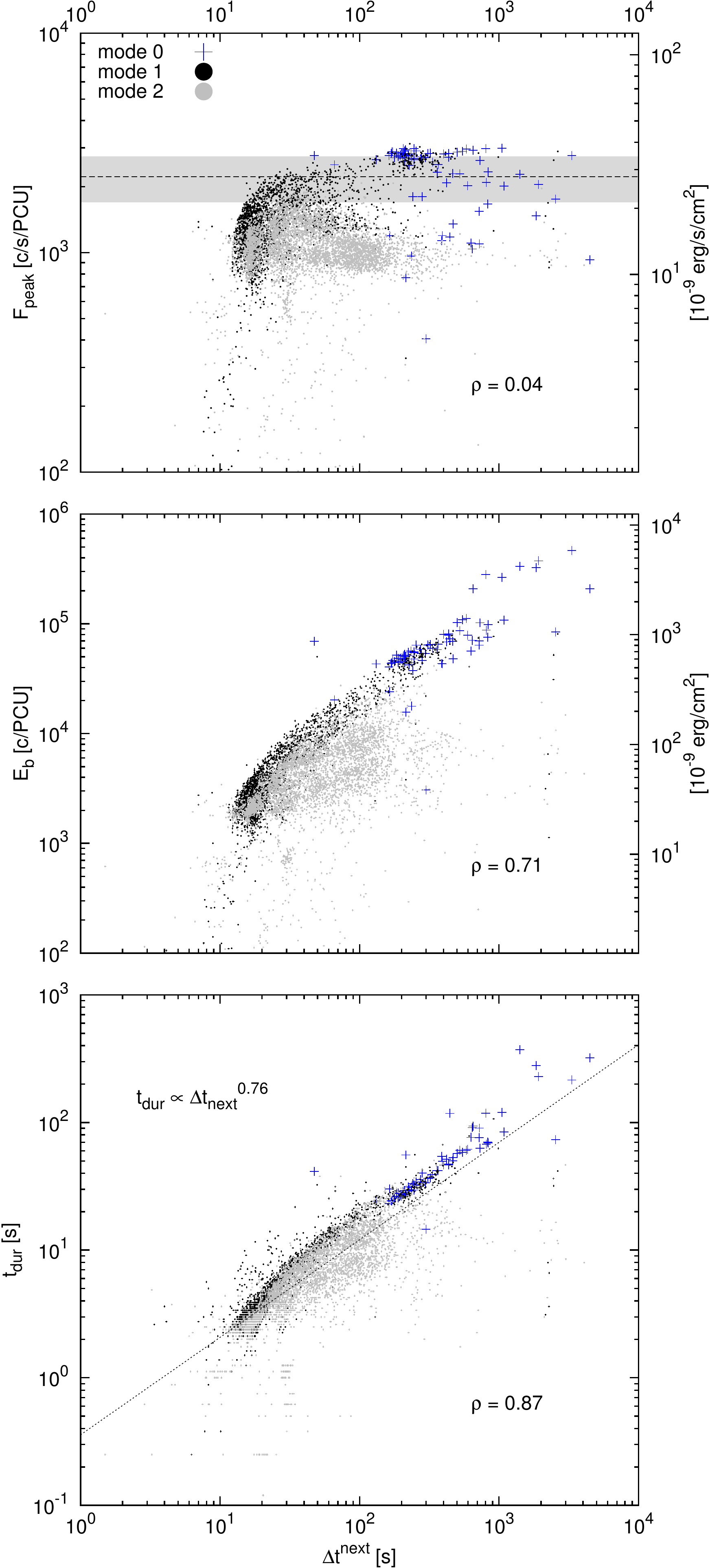}
  		\caption{\small{Burst peak flux $F\sub{peak}$, fluence $E\sub{b}$,
		duration $t\sub{dur}$ as a function of the
		waiting time $\Delta t$ to the next burst (see Sec.~\ref{sec:typeIIbursts}).
		Count rates and energies are as in Fig.~\ref{fig:outburst92026}.
		The horizontal dashed line indicates the Eddington luminosity,
		with the grey-shaded area indicating the uncertainties on the distance and
		the spectral fit (see Sec.~\ref{sec:proxy}).
		The middle and bottom plot show the relaxation oscillator behaviour of the type II bursts;
		the burst recurrence time seems to more accurately predicted
		by the duration of the previous burst than its fluence.
		The least-squares fit to the data in the bottom plot is also shown 
		with a dotted line (see Sec.~\ref{sec:typeIIbursts}).
		}}
	\label{fig:burst_rels2}
\end{figure}

\begin{figure}
  	\includegraphics[width=\columnwidth]{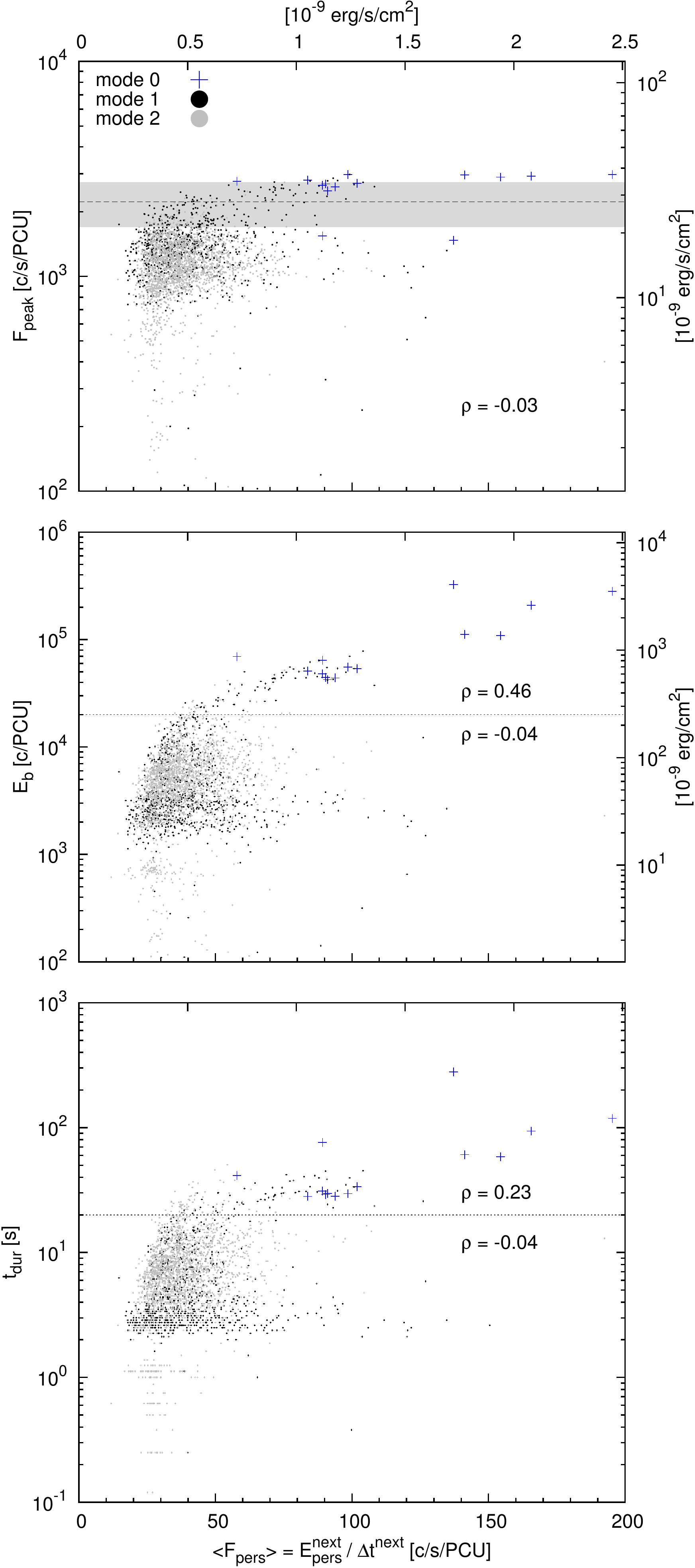}
  		\caption{\small{Same burst properties as in Fig.~\ref{fig:burst_rels2},
		here plotted against the average persistent flux $\langle F\sub{pers}\rangle$
		(see Sec.~\ref{sec:typeIIbursts} for definition).
		Only bursts observations that were not contaminated by 4U 1728-34
		are plotted (see Sec.~\ref{sec:sample}).
		Count rates and energies are as in Fig.~\ref{fig:outburst92026}.
		The horizontal dashed line indicates the Eddington luminosity,
		with the grey-shaded area indicating the uncertainties on the distance and
		the spectral fit (see Sec.~\ref{sec:proxy}).
		No overall correlation is visible in the data,
		although it might hold for the longest,
		most energetic bursts (see Sec.~\ref{sec:typeIIbursts}).}}
	\label{fig:burst_rels3}
\end{figure}

\begin{figure}
  	\includegraphics[width=\columnwidth]{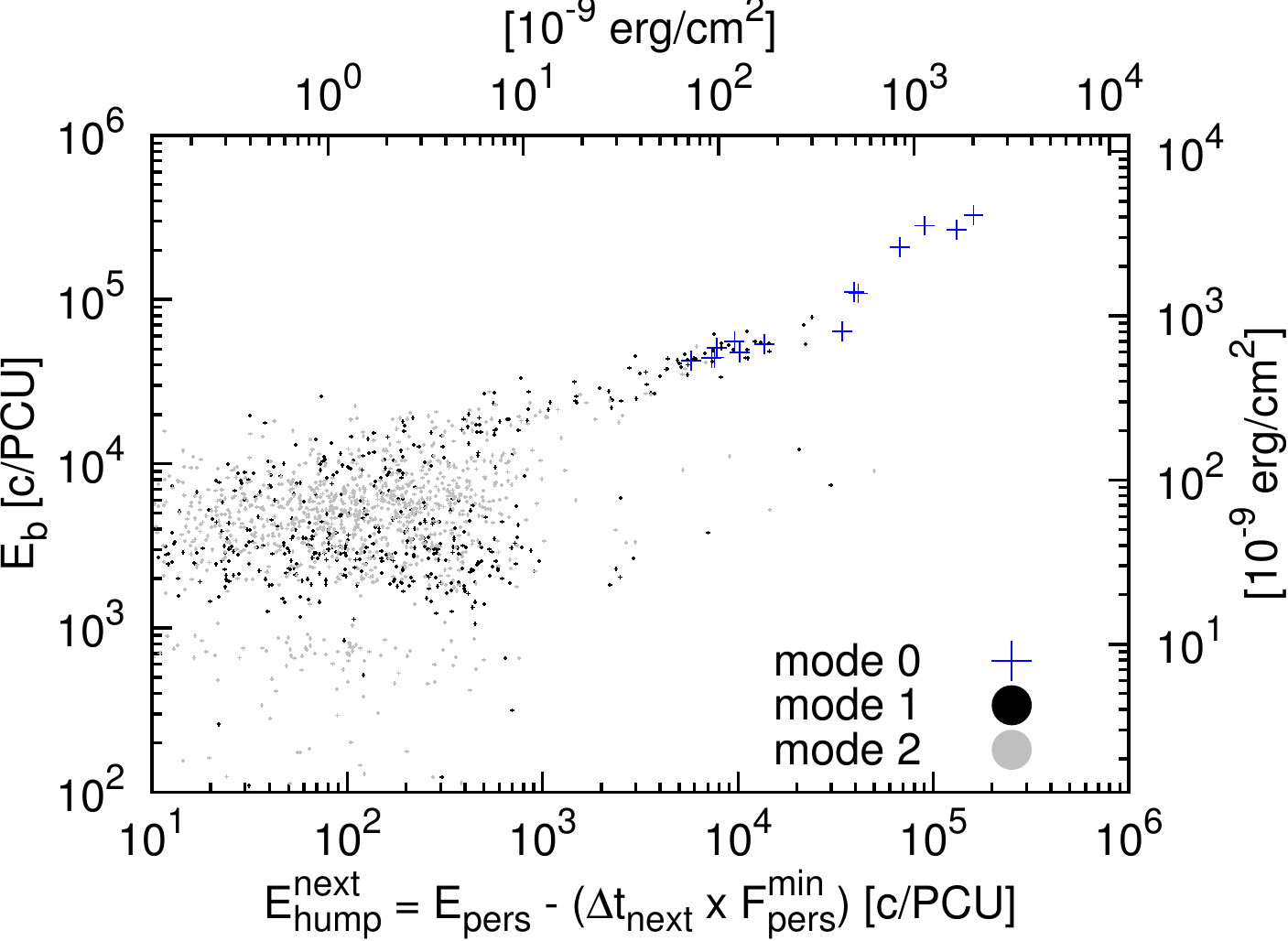}
  		\caption{\small{The burst fluence $E\sub{b}$
		against the ``net'' (i.e. above $F\sub{pers}\sur{min}$)
		hump fluence $E\sub{hump}$ (see Sec.~\ref{sec:typeIIbursts}).
		Count rates and energies are as in Fig.~\ref{fig:outburst92026}.
		Above a critical $E\sub{b} \sim 2 \times 10^4$~c/PCU,
		a correlation appears.
		Both the brightest mode 1 bursts and all mode 0 bursts
		can be above this threshold.}}
	\label{fig:burst_rels4}
\end{figure}

Having described the burst properties individually,
we turn our attention to
the relationships between them.
As we show in Fig.~\ref{fig:burst_rels1},
some burst properties show trivial correlations.
The burst fluence $E\sub{b}$ correlates with both the peak flux $F\sub{peak}$
(Spearman rank-order correlation coefficient $\rho = 0.57$)
and, more strongly, with the burst duration $t\sub{dur}$
($\rho = 0.90$), with the mode 0 bursts
having the largest $E\sub{b}$ and longest $t\sub{dur}$.
In the lower left corner of the upper plot
are the bursts for which almost the entire fluence is emitted
in a short interval around the burst peak
(in the lower plot these bursts are mostly shorter than 10~s).
At the opposite end of the top plot,
the correlation between $F\sub{peak}$
and $E\sub{b}$ saturates once the peak flux reaches the Eddington limited.
There seems to be no correlation between
peak flux and duration (not plotted).

It is also interesting to see how these quantities,
which determine the burst energetics,
depend on the bursting rate at a given moment.
Fig.~\ref{fig:burst_rels2} shows
the type II burst peak flux, fluence and duration as a function
of the recurrence time to the next burst $\Delta t\sur{next}$.
On top, one can see that $F\sub{peak}$
is roughly independent of the waiting time ($\rho = 0.04$).
A correlation instead appears in the middle panel,
which plots $E\sub{b}$ against $\Delta t\sur{next}$ ($\rho = 0.71$):
this is the well-known relaxation-oscillator behaviour
of the RB, according to which the more energetic a burst,
the longer it takes for the next one to occur.
This is the reason why we chose
to introduce the parameter $\alpha\sur{next}$.
However, a look at the bottom plot in Fig.~\ref{fig:burst_rels2}
reveals that $\Delta t\sur{next}$ correlates more tightly with
$t\sub{dur}$ ($\rho = 0.87$) than with $E\sub{b}$.
This implies that the duration of a type II burst,
rather then the energy it releases, determine
the time it takes for the next type II burst to occur.
A least-squares fit yielded $t\sub{rec} \propto (\Delta t\sur{next})^{0.76 \pm 0.01}$.
There is, however, a large spread in the data,
especially for the mode 1 and mode 2 bursts.

As we explained in Sec.~\ref{sec:outburst}
different patterns of type II bursts follow each other during the outburst decay,
while on average the persistent flux decreases.
We therefore wish to assess whether the burst properties
show any dependence on the strength of the persistent emission.
Fig.~\ref{fig:burst_rels3} shows the burst peak flux, fluence and duration
plotted against the average persistent flux $\langle F\sub{pers}\rangle$.
It is immediately evident that
all correlations are weaker than with respect to $\Delta t\sur{next}$.
As in the previous case, $F\sub{peak}$ shows no evident relation
to the persistent emission properties ($\rho = -0.03$).
In the middle and bottom panels,
where we plot $E\sub{b}$ and $t\sub{dur}$ against $\langle F\sub{pers}\rangle$,
no monotonic trends can be seen.
Instead, two distinct behaviours appear.
While for bursts that liberate little fluence ($E\sub{b} \lesssim 2\times 10^4$~c/PCU)
or that are short ($t\sub{dur} \lesssim 20$~s) 
$E\sub{b}$ and $t\sub{dur}$ do not correlate with
$\langle F\sub{pers}\rangle$ ($\rho = -0.04$ in both cases),
for energetic and long bursts a positive correlation is visible,
that at least for $E\sub{b}$ appears to be significant ($\rho = 0.46$ and 0.23 respectively).

In other words, when looking at how the burst properties
are influenced by the persistent emission,
there is a dichotomy in behaviour between the
mode-0 type II bursts, which have
large fluences and long durations, and appear in the upper half of the plots,
and mode 2 bursts, characterized by small fluences and short durations,
appearing in the bottom half.
Mode 1 bursts can clearly be found both in the
lower half of the plots of Fig.~\ref{fig:burst_rels3} (low fluence and short duration)
and in the upper half (large fluence and long duration).
These are, respectively, the short bursts initiating the mode 1 sequence,
and the long one that concludes it.
These two groups inhabit the same regions of the plots
as the mode-2 and 0 bursts respectively.

The same threshold in fluence seems to determine
the presence of a hump after a burst.
Fig.~\ref{fig:burst_rels4} shows that,
independent of burst mode,
above a critical $E\sub{b} \sim 2\times$\ergcm{-7}, the burst fluence
correlates with the ``net'' (i.e. above $F\sub{pers}\sur{min}$)
hump fluence $E\sub{hump}$.
Again, bursts above this threshold are not only the mode 0,
but also the resembling mode 1 that end a sequence of bursts and precede a hump.

\subsection{Type I bursts in the mixed (hard) state}\label{sec:typeIhard}

\begin{figure}
  	\includegraphics[width=\columnwidth]{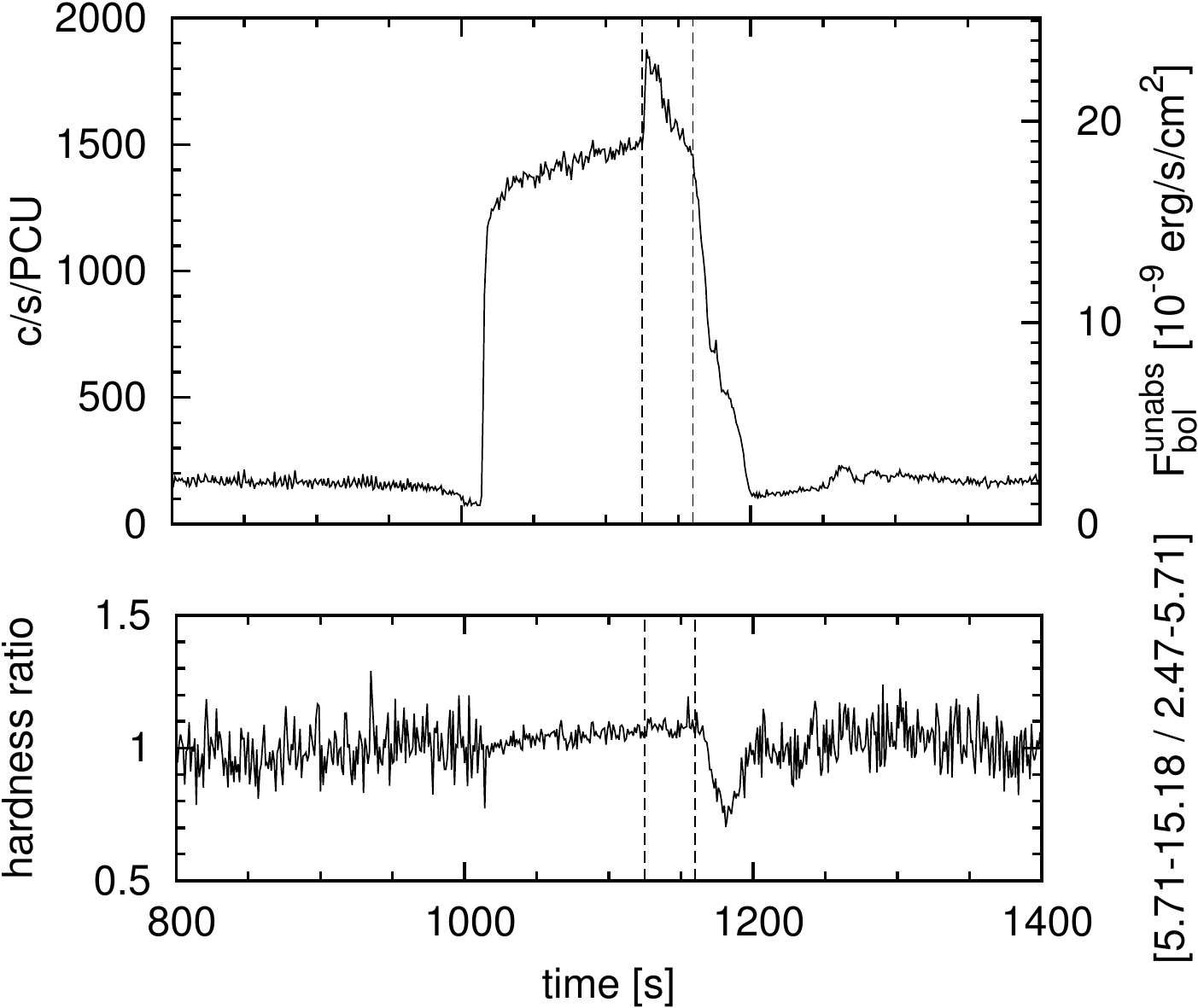}
  		\caption{\small{Light curve and hardness ratio of ObsID 50420-01-09-00R,
		featuring a type I burst taking place during a mode 0 type II burst.
		The vertical lines indicate the approximate start and end time of the burst.
		The net peak flux to persistent flux ratio $\beta \lesssim 0.2$,
		hiding the hardening of the flux during the type I burst
		below the detection level \citep[][B13]{2011ApJ...733L..17L}.}}
	\label{fig:typeImode0}
\end{figure}

\begin{figure}
  	\includegraphics[width=\columnwidth]{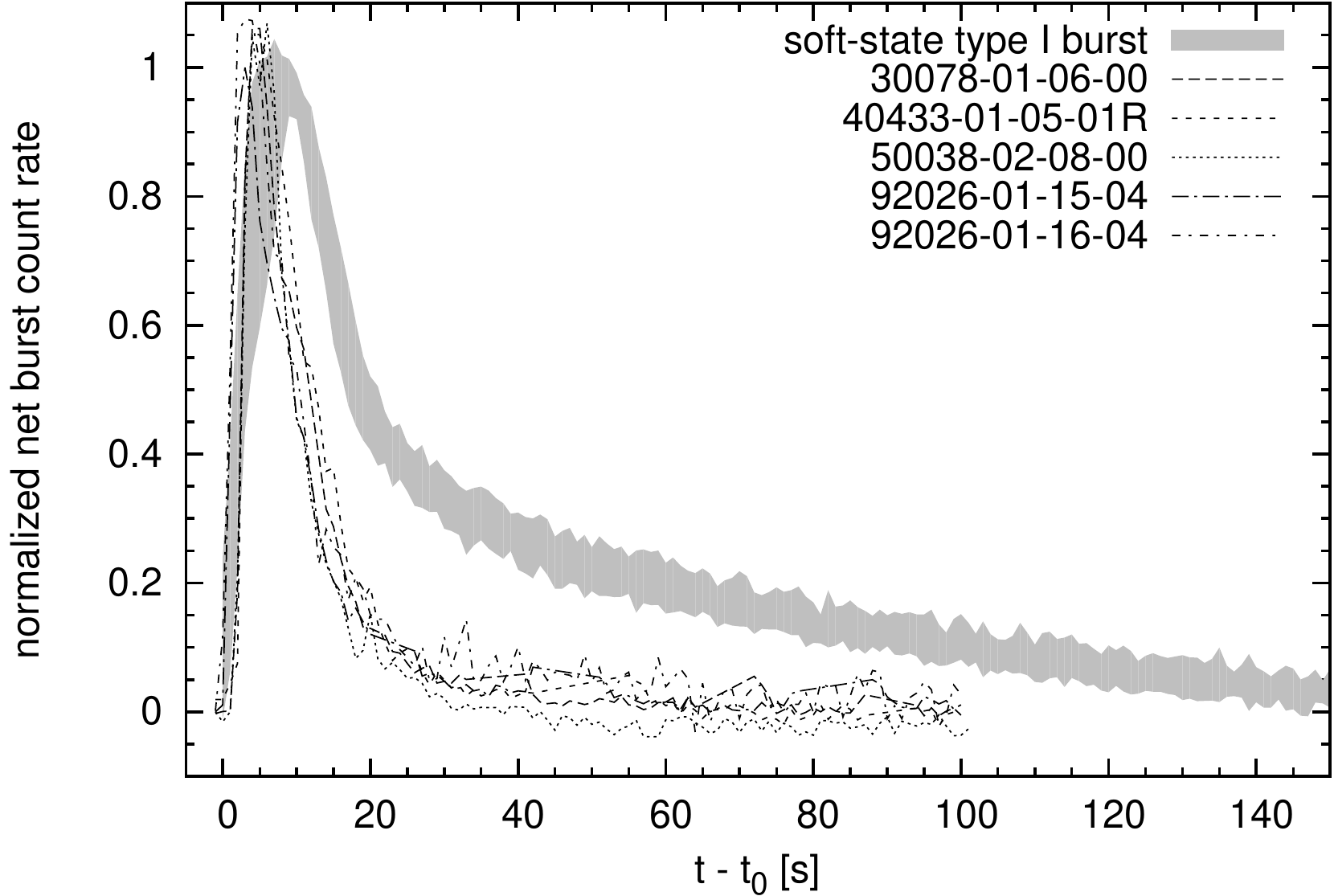}
  		\caption{\small{Light curves of 5 type I bursts in the hard state
		plotted against the average profile of the 22 single-peaked type I bursts
		featuring the longest and brightest decays (see B14),
		all of which are instead observed in the soft state.
		The net bursts counts have been normalized so that
		the average flux during the burst peak
		(which is defined as the interval during which the flux is above 90 per cent
		of the burst peak flux) is 1.
		The vertical extent of the shaded region indicates
		the standard deviation of the soft-state sample.
		The hard-state type I bursts were randomly chosen among those
		taking place on top of a stable persistent emission,
		i.e. outside humps, and showing no overlapping type II bursts during the
		first 40~s of the type I burst.}}
	\label{fig:typeIhard}
\end{figure}

\begin{figure}
  	\includegraphics[width=\columnwidth]{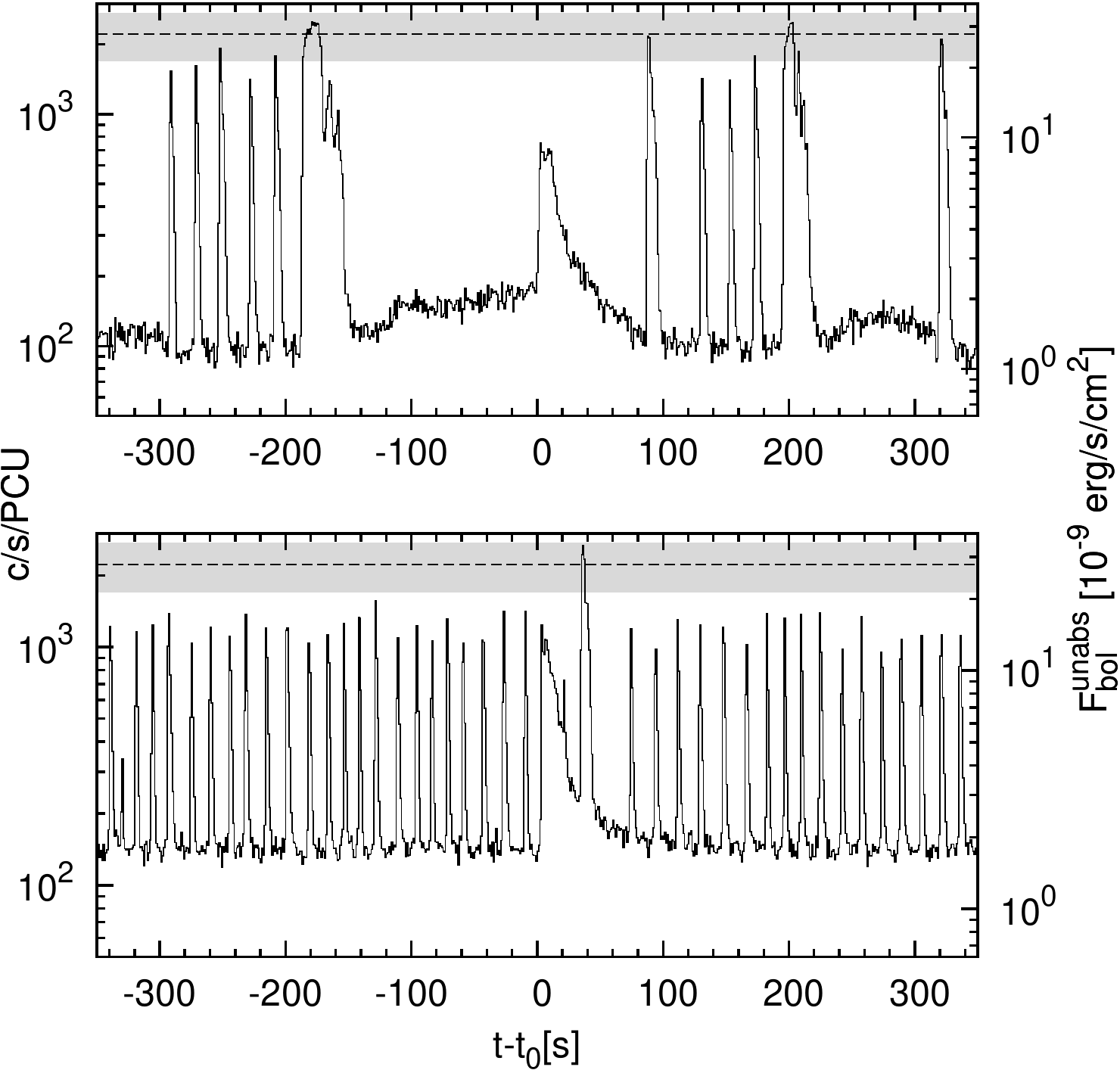}
  		\caption{\small{Two examples of type I bursts during the hard state,
  		during mode 1 (\textit{top}) and 2 (\textit{bottom}) type II bursts respectively.
  		The $x$ axis shows time with respect to the type-I-burst
  		start time $t_0$.
		The y-axes are as in Fig.~\ref{fig:outburst92026}.
		The horizontal dashed line indicates the Eddington luminosity,
		with the grey-shaded area indicating the uncertainties on the distance and
		the spectral fit (see Sec.~\ref{sec:proxy}).
  		Notice that the type I burst during the mode 1 phase takes place
  		during the hiatus in type II bursts that follows a long one,
  		while the persistent emission level rises in a hump shape.
		This is the case for 10 out of 12 type I burst during the mode 1 phase
  		Neither effect is observable as a type I burst takes place in the mode 2 phase
		(see Sec.~\ref{sec:typeIhard}).
		The type I burst during the mode 2 phase briefly affects the type II burst emission.
		This is only observed in 1 case out of 11.
  		}}
	\label{fig:tI_1vs2}
\end{figure}

Various methods help identify the type I bursts in the sample.
Their profiles in the light curves
differ for a number of features from those of type II bursts.
The type-I burst peak fluxes are smaller than those of most type II bursts
(type I bursts in the RB reach at most one third of the Eddington luminosity, B13).
Also, all type II bursts show multiple peaks in their decays,
while type I bursts generally have smooth decays (but see B14).
Finally, type I bursts have generally longer durations,
causing the fast-recurring type II bursts to overlap with their exponential decays.
Finding candidate type I bursts is therefore relatively straightforward.

Nonetheless, a secure identification of type I bursts
should come from measurements
of $\alpha$ or from the detection of cooling in their time-resolved spectra.
However, the first method is only applicable in a state where \textit{either}
type I or II bursts are emitted. In a mixed state,
where type II bursts recur faster than the type I bursts,
one cannot rely on their 
$\alpha\sub{prev}$ values
to discriminate between them.
The second method, the detection of cooling, is only a sufficient,
and not necessary condition for the type I burst identification.
Observing spectral cooling is generally only possible
for bright type I bursts \citep[$F\sub{peak} \gtrsim 0.7 F\sub{pers}$;][]{2011ApJ...733L..17L}.
B13 also showed that some RB bursts which lack evidence for cooling are
to be identified as type I,
due to their large $\alpha$ values and their occurrence
in the brightest phases of the soft state.

To determine the presence of cooling,
we measured hardness ratios as the
flux ratios in the 5.1-7.6~keV and 2.3-5.1~keV bands.
26 candidate bursts show secure evidence of cooling,
and we can therefore confirm they are type I even without
being able to measure their $\alpha$ values.

Additionally, we found three bursts taking place
``on top'' of a mode 0 type II bursts
(see Fig.~\ref{fig:typeImode0}).
For these bursts $F\sub{peak} \lesssim 0.2 F\sub{pers}$,
meaning the lack of cooling in their spectra cannot rule out their type I nature.
Given their distinctive shape (fast rise and exponential decay),
these three bursts are unlikely to be a sudden spike
of the type II bursts, which tend to show
nearly flat plateaus (see Sec.~\ref{sec:outburst}).
Therefore we believe they are type I bursts as well.

Summarizing, we found a total of 29 type I bursts occurring in an observation
where also type II bursts appeared,
out of a total of 123.
Note that 17 candidate type I bursts could not be confirmed.

As shown in Fig.~\ref{fig:typeIhard},
the type I bursts in the hard state are
significantly different from those in the soft state.
Most remarkably, the durations are much shorter, with $t\sub{dur}\lesssim$40~s,
as are the peaks, that lack the 6-8~s plateau
displayed by type I bursts in the soft state (B13).

Out of 29 type I bursts identified in the hard state,
6 took place around mode 0 type II bursts,
12 during mode 1, and 11 during mode 2.
Ten out of 12 type I bursts found during mode 1 activity
occur during the intervals of enhanced persistent emission (``humps'')
that are observed between two more tightly packed sequences of type II bursts
(see Fig.\ref{fig:tI_1vs2}, top).
Only two type I bursts are instead observed ``inside'' a sequence of mode 1 type II bursts.
No such preferential location is observed for the 11 type I bursts
occurring during the mode 2 phase, when the persistent emission count rate
and the recurrence time of the type I bursts are roughly constant.
However, it must be noted that one such type I burst seems to briefly ($<50$~s)
affect the type II emission
(see Fig.\ref{fig:tI_1vs2}, bottom), 
which is otherwise stable in this observation,
as is common in mode 2.

For mode-1 type-II bursts, one can calculate the chance
of random coincidence of a hump and a type I burst.
We examined all the mode-1-burst sequences
during the outburst plotted in Fig.~\ref{fig:outburst92026},
and summed the integrated fluence from
the beginning of each burst in a sequence until the last, longer one.
We divided this integrated energy by the fluence
emitted during the following hump.
Over the course of the outburst, this ratio decreases
from 13.5 to 2.7.
In order to reach a conservative estimate,
we take the latter as the ratio of the amount
of mass accreted during a mode-1-burst sequence
to that accreted during a hump.
This means that, for each individual burst,
the probability of random occurrence
of a burst on a hump is $p\sub{single} = 0.27$.
An estimate of the probability
that 10 out of 12 type I bursts
randomly take place on a hump rather than overlap with mode 1 bursts is
thus given by the binomial distribution as $p_{10/12} = 7.24 \times 10^{-5}$.
Clearly type I bursts have a statistically significant
preference for occurring during breaks in the type II activity,
rather than overlapping them.

B13 reported type I burst recurrence times
in the soft state in the range 0.4--4~ks.
As the RB enters the hard state, $t\sub{rec}$
becomes difficult to measure,
as it typically surpasses the duration of an observation.
The numbers should therefore only be treated as upper limits.

Nonetheless, we can establish  average recurrence times
of type I bursts across the different modes of type II bursting.
There are 6 confidently identified 
type I bursts in 114~ks of observations of mode 0,
12 in 170~ks of mode 1,
and 11 in 492 ks of mode 2.
If one includes the dubious cases,
the type I bursts are 13 in mode 0,
21 in mode 1 and 12 in mode 2.
This gives upper and lower limits, respectively,
on the average recurrence times:
$\langle t\sub{rec}\sur{mode 0} \rangle \sim (8.3-18.1)$~ks,
$\langle t\sub{rec}\sur{mode 1} \rangle \sim (8.1-14.1)$~ks
and $\langle t\sub{rec}\sur{mode 2} \rangle \sim (41.0-44.7)$~ks.
Thus, the type I burst recurrence time is roughly
the same during mode 0 and mode 1.

We plot these recurrence times in Fig.~\ref{fig:trec}
together with the type I burst recurrence times
measured in the soft state,
against the persistent flux $F\sub{pers}$ (B13).
The latter was taken to be a proxy of the mass accretion rate
on the source, and the data show a correlation,
with $t\sub{rec} \propto F\sub{pers}^{-0.95}$.
The $\langle F\sub{pers} \rangle$ parameter
we define in Sec.~\ref{sec:variables}
only measures the emission taking place between type II bursts,
but these should also accrete nuclear fuel onto the surface.
Therefore, we plot the type I burst recurrence times
across the three modes
versus both $\langle F\sub{pers} \rangle$ and $F\sub{tot}$,
which includes the burst fluence in the average.

As Fig.~\ref{fig:trec} shows,
regardless of the proxy that is chosen
for the mass accretion rate,
the largest recurrence time for type I bursts is
observed during mode 2 type II bursts.
Interestingly, observations
line up with the relation extrapolated from the soft state
when employing $\langle F\sub{pers} \rangle$,
while for $F\sub{tot}$
there appears to be a scarcity of type I bursts in the hard state,
across all modes.

\section{Observational summary}\label{sec:sum}

The large sample of data we have studied allows us to address
fundamental questions over the type II bursts in the RB.
In this section we first summarize the overall properties of the sample,
starting with those that were already previously known:

\newcounter{bean}
\begin{list}{\arabic{bean})}{\usecounter{bean} \itemindent=-0.5cm \leftmargin=0.5cm \itemsep=0cm \parsep=0cm \topsep=0cm}
 \item There are four patterns of type II bursts (e.g., Fig.~\ref{fig:outburst92026}).

\newcounter{subbean}
\begin{list}{\roman{subbean})}{\usecounter{subbean} \itemindent=-0.5cm \leftmargin=0.4cm \itemsep=0cm \parsep=0cm \topsep=0cm}
 \item a (type II) burst-less soft state that is visible
when the persistent emission flux is above 10 per cent of the Eddington limit.
This occurs mostly at the beginning of outbursts.

\item During the outburst decay, the first type II bursts to appear
are of so-called mode 0 \citep{1999MNRAS.307..179G}.
They have the longest durations and recurrence times, and the largest fluences.
The flat tops often observed in long mode 0 bursts
are often sub-Eddington.
Indeed, the flatness of the bursts has before been noticed
to be a property of long bursts \citep{1991MNRAS.251....1T},
regardless of their luminosity.

\item The mode 1 burst pattern follows,
with sequences of frequent short bursts ending with a longer and brighter one
that is similar in appearance to the shortest mode 0 bursts.

\item Finally, the mode 2 pattern appears,
consisting of short bursts like the short ones in mode 1
(but not interrupted by larger bursts followed by longer recurrence times).
\end{list}

This is only a general rule and exceptions exist.
For instance, the RB sometimes briefly switches back to the previous mode,
and even back into the soft state.
Mode 0 and 1 bursts can last for up to $\sim 5-6$ days,
while mode 2 bursts persist for up to 24 days.

 \item The persistent emission decreases monotonically
as the RB progresses from mode 0 to mode 1 to mode 2 bursts (Fig.~\ref{fig:persistent_trend}).
No type II bursts are observed on top of undetectable persistent emission.

 \item All bursts roughly follow a relaxation oscillator relation
(Fig.~\ref{fig:burst_rels2}, middle),
meaning that the fluence in a burst determines the waiting time to the next burst.

 \item 95 per cent of the bursts show $\alpha$ values below 1.5,
one order of magnitude below the minimum value
that is possible for full thermonuclear burning.
The distribution peaks at 0.2,
meaning that 5 times more fluence is liberated in the
bursts than in the persistent emission.
This underlines the nature of type II bursts as accretion processes.

\item We calculated the probability of random occurrence of
the observed fraction of type I bursts on a hump
to be $p_{10/12} = 7.24 \times 10^{-5}$.
Clearly, type I bursts during mode 1
have a statistically significant
preference for occurring during breaks in the type II activity,
rather than overlapping them (see Fig.\ref{fig:tI_1vs2}).
This preference was first noticed in SAS-3 data by \citet{1977ApJ...214L..11U},
although it was \citet{1978Natur.271..630H} who first proposed that
the bursts on the humps and those surrounding them were type I and II respectively.
They also noticed that these bursts were ``anomalous''
with respect to the type I bursts taking place during what we call mode 2,
in that they had smaller peak fluxes and sometimes lacked spectral softening during burst decay.
Likewise, we could confirm the presence of spectral softening
in 11 out of 12 candidate type I bursts during mode 2,
but only in 12 out of 21 during mode 1.

\end{list}
New findings in RB type II bursts are:

\begin{list}{\arabic{bean})}{\usecounter{bean} \itemindent=-0.5cm \leftmargin=0.5cm \itemsep=0cm \parsep=0cm \topsep=0cm}
\setcounter{bean}{5}
 \item The range of minima in the persistent emission between bursts $F\sub{pers}\sur{min}$
is between 1 and 2.5 per cent of the Eddington luminosity.
The distribution of the actual average persistent fluxes (i.e. taking the humps into account)
$\langle F\sub{pers}\rangle$ extends to only slightly larger values,
because the general sample is overwhelmingly dominated by the short-mode-1 and mode 2 bursts,
without humps in between.
Taking into account the burst fluence as well,
the RB is on average accreting at up to $\simeq 0.17 L\sub{Edd}$
(Fig.~\ref{fig:persistent_hist}).

\item The persistent flux does not generally seem to correlate
with the burst properties (Fig.~\ref{fig:burst_rels3}).
Each burst mode presents a large dispersion in the range of observed
persistent fluxes (Fig.~\ref{fig:persistent_trend}),
making the latter a poor predictor of the burst mode.

 \item The most energetic type II bursts reach the Eddington limit
($F\sub{Edd} = 2.8\times$\ergscm{-8};
Fig.~\ref{fig:burst_hist}, top-left).
1 per cent of the bursts are super Eddington.
However, since the maximum $F\sub{peak}\sur{max} = 3.0\times$\ergscm{-8},
even a slight decrease in the H abundance
or moderate beaming are enough to bring
every single burst within the uncertainty on $L\sub{Edd}$ (Sec.~\ref{sec:proxy}).
Bursts reaching the Eddington limit
are not accompanied by evidence for 
photospheric radius expansion in the time-resolved spectra.
Generally speaking, the peak flux shows no correlation with
the waiting time or the persistent flux properties,
and the distribution of peak fluxes is relatively narrow.
Even the only visible correlation,
the one with the burst fluence (Fig.~\ref{fig:burst_rels1}),
shows a large spread.
This suggests that the speed at which energy is released
does not depend on the
total amount of energy available for a burst,
nor on the global mass accretion rate.

 \item There is a clear dichotomy between mode 0 and mode 2 bursts.
The former appear to feed back into the persistent emission,
which rises in between them, while the latter
do not seem to affect it.
This is also reflected in the fact that only for mode 0 bursts
there appear to be correlations between some properties of the bursts
and of the persistent emission (Figs.~\ref{fig:burst_rels2}, \ref{fig:burst_rels3}).
Mode 1 bursts clearly are an intermediate state of the instability,
reproducing both behaviours.

 \item Long bursts -- meaning both mode 0 bursts and the mode 1 bursts that ``close''
a sequence of bursts -- are followed by humps, i.e. burst-free enhanced persistent emission.
This is partially contrary to what was previously believed, that the persistent emission
``dips'' between mode 0 bursts
\citep{1979ApJ...227..555M,1984PASJ...36..215K,1988ApJ...324..379S,1992MNRAS.258..759L,1993SSRv...62..223L,1999MNRAS.307..179G},
despite the fact that there is no evidence
for enhanced absorption in the spectra \citep{1988ApJ...324..379S,1993MNRAS.261..149L}.
However, looking at periods when the RB switches
between these modes of emission, we clearly show that such ``dips''
are actually at the same minimum level observed during a sequence of mode 1 bursts
(Fig.~\ref{fig:mode_switching}).
Since the persistent emission level is stable inside a mode 1 sequence,
and no increase in the column density $N\sub{H}$ is observed in the spectra
of the ``dips'' (see Appendix~\ref{sec:spectralstudy}),
we argue that in both cases the emission following a long burst is enhanced,
to decay again in the time running to the next burst.

 \item The recurrence time distribution of type II bursts
is strongly peaked at $\Delta t = 15-18$~s,
and drops abruptly below this value, while
showing a smoother decrease to larger recurrence times (Fig.~\ref{fig:burst_hist}).
Bursts with shorter recurrence times are found (as short as 2~s)
but they are a much smaller fraction of the sample,
despite the fact that for a shorter $\Delta t$
more bursts should be produced.
This implies that a minimum time exists for the instability
to develop, and that it is difficult for the RB to
produce type II bursts more quickly than that.

 \item We found 61 bursts with durations shorter than 1~s,
and as short as 130~ms (Fig~\ref{fig:micro}, top),
one order of magnitude smaller than the shortest bursts
so far reported for type II bursts
\citep[2~s in \textit{EXOSAT} data, ][]{1991MNRAS.252..190L}.
This puts a constraint on theoretical models for this instability,
that appears to cover four orders of magnitude in duration,
up to the record observed $t\sub{dur} = 1100$~s.

\item The sub-s duration bursts have $\alpha$ values between 20 and 200,
thus overlapping with the $\alpha$ range of type I burst.
Precursor events of sub-s durations have been observed before,
but they are always followed by long type I bursts
\citep{2014A&A...568A..69I}.
Therefore, the sub-s bursts are of type II.

\item The relaxation-oscillator behaviour 
is more accurately predicted
by the burst duration than the fluence (Fig.~\ref{fig:burst_rels2}, bottom).

\end{list}

\section{Discussion}\label{sec:dis}

Having laid out all observational features of the type II bursts of the RB,
we now discuss the physical implications of these.
We leave the discussion of the instability models for type II bursts
to the next Section.

\subsection{Accretion versus ejection}

Two observables might suggest the possibility
of mass ejection in the RB.
One comes from the type I burst behaviour,
which we discuss in Sec.~\ref{sec:trectypeI}.
The second finding suggesting that mass could be ejected
during type II bursts is that
their peak flux never,
for any persistent flux or burst fluence,
surpasses a certain value which,
within a factor of 2, is equal to the expected Eddington limit
for a NS at the distance of the hosting globular cluster,
with $X=0.7$ and an isotropic radiation field
(Figs.~\ref{fig:burst_rels1} through \ref{fig:burst_rels3}).
One might therefore wonder whether any surplus
radiation power is transformed to
the kinetic energy of an outflow enforced by radiation pressure.

There is, however, no spectral evidence for expansion
of the emitting region.
Thus, the picture arises that
when the Eddington luminosity is reached, the accretion is temporarily halted,
interrupting the release of gravitational energy.
This in turn lowers the luminosity so that accretion can ensue again,
rising back to the Eddington limit, and so on.
The radiation pressure may momentarily cut off the fuel line for the radiation
(the accretion flow), but not result in isotropic expansion.
As a consequence of this, 
all energy is liberated through radiation,
and there are no kinetic losses.
This is in contrast to the type I bursts,
where reaching the Eddington luminosity
does not affect the energy release from the layer that
has been heated by the burning, and the photosphere expands.
Instead, once a type II burst flux reaches the Eddington luminosity,
a larger fluence $E\sub{b}$
can only be liberated by means of an increased duration,
which makes for a tight relation between burst duration and fluence
(Fig.~\ref{fig:burst_rels1}, bottom).

\subsection{Two kinds of type II bursts and three kind of modes}

The data point to two kinds of type II bursts:
long, often luminous ones, and short ones.
Long bursts, predominantly mode 0,
occur at relatively larger mean accretion rates,
while shorter ones abound at lower rates, during mode 2.
They mix in mode 1. Long bursts induce humps
and adhere more tightly to a relaxation oscillator behaviour,
while the short ones are not followed by humps
and show a larger scatter around the relaxation oscillator relation.

The evidence for this is provided in Fig.~\ref{fig:burst_rels3}.
At large $\langle F\sub{pers}\rangle$ one observes
both long and energetic bursts,
that become longer and stronger at larger persistent fluxes,
and short and weak ones, displaying an opposite trend
(i.e. becoming shorter and weaker at larger $\langle F\sub{pers}\rangle$).
The critical fluence separating
these two opposite feedbacks is $E\sub{b} \approx 2 \times$\ergs{39}.
This duality is reflected by the way in which the bursts
affect the persistent emission back.
While short, weak burst do not have a feedback on $F\sub{pers}$,
the long and strong bursts that appear in mode 0 and at the end of a mode 1 sequence
generate humps of enhanced persistent emission.
Fig.~\ref{fig:burst_rels4} shows that the threshold for this behaviour
is again at $E\sub{b} \approx 2 \times$\ergs{39}.

Another look at Fig.~\ref{fig:burst_rels1}
shows that the threshold in fluence at which the bursts become Eddington limited
is $E\sub{b} \sim 5 \times$\ergs{38}.
Notice that some mode 2 type II bursts reach $F\sub{peak} = F\sub{Edd}$,
but very quickly decay, and are not followed by humps.

Possibly then, a sustained period of Eddington-limited emission
is necessary for a type II burst to be followed
by a period of enhanced persistent emission.
It would therefore make more sense to speak of ``short'' and ``long'' bursts,
rather than the three modes that are mentioned in literature.
What Fig.~\ref{fig:burst_rels4} also shows, however, is that
there is a continuity in behaviour between the two, rather than a clear cut
distinguishing two different behaviours among type II bursts.

\subsection{The relaxation-oscillator behaviour}

Our data confirm the existence of the well-known relaxation oscillator
between the burst fluence $E\sub{b}$ and the time to the next burst $\Delta t\sur{next}$.
(Fig.~\ref{fig:burst_rels2}, middle).
However, we also find a tighter correlation between 
the burst duration $t\sub{dur}$ and $\Delta t$
(Fig.~\ref{fig:burst_rels2}, bottom).
It is quite possible that the relaxation-oscillator relation
is a by-product of this tighter correlation and of
the aforementioned one between $E\sub{b}$ and longer $t\sub{dur}$.
This would mean that it is the timescales, rather than the energetics,
that connect the bursts and the persistent emission.

The burst properties seem to depend weakly
on the average intra-burst persistent flux (Fig.~\ref{fig:burst_rels3}).
Also, relatively small changes in the persistent flux are visible between modes.
The variations in the burst properties are therefore much larger
than those in the accompanying persistent emission.
In other words, the variation in the global mass accretion rate
that must be present between the beginning and the end of the hard state
is not equally split between the bursts and the persistent emission,
but almost entirely goes into the former.

\subsection{Type I bursts as a probe of mass accretion by type II bursts}\label{sec:trectypeI}

\begin{figure}
		\includegraphics[width=\columnwidth]{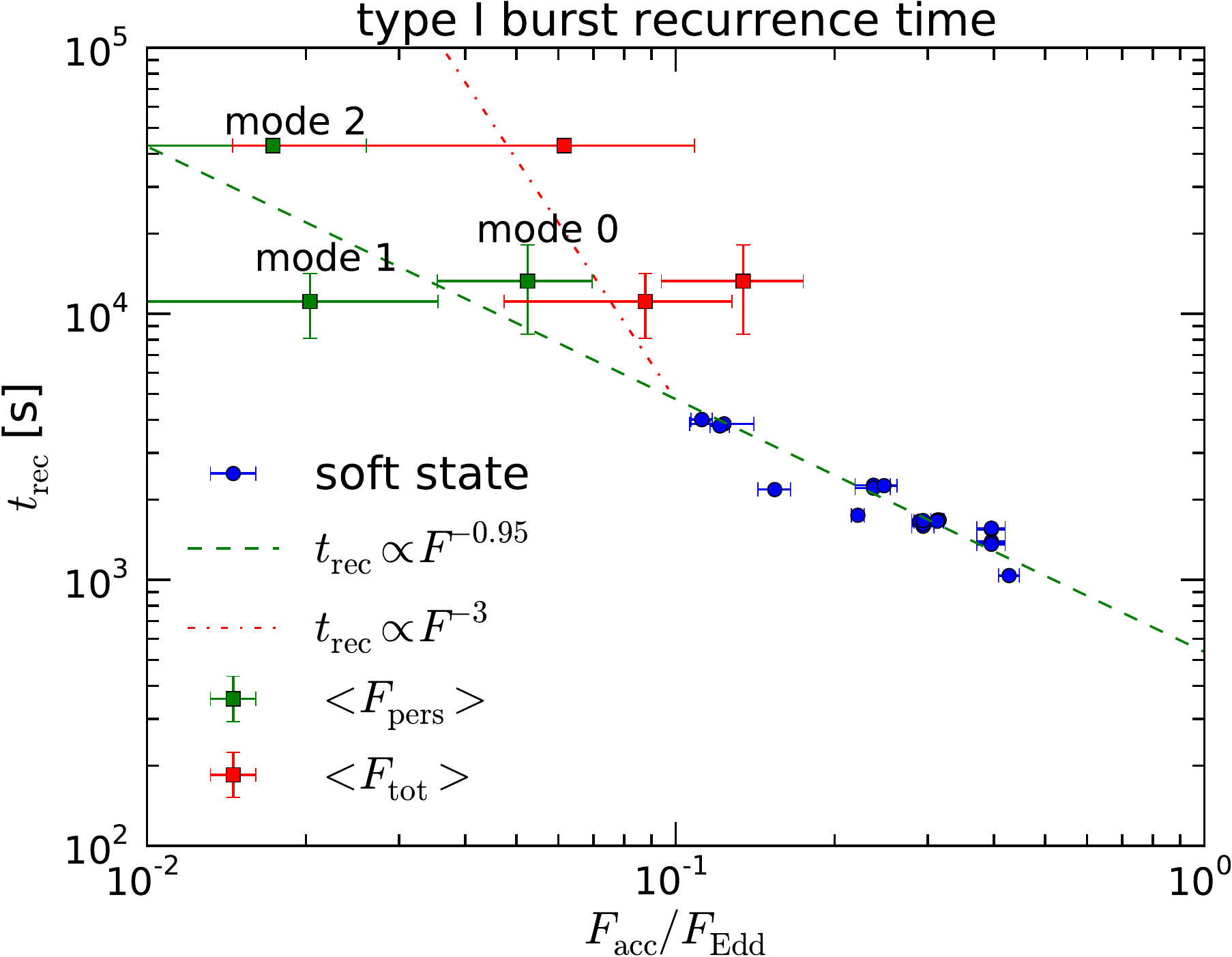}
			\caption{\small{The recurrence time $t\sub{rec}$ of type I bursts
			at different stages in the outburst.
			Blue points are soft-state measurements from B13,
			from which the fit is derived $t\sub{rec} \propto F\sub{pers}^{-0.95}$
			(\textit{dashed line}).
			Squares are averages
			for $t\sub{rec}$ during mode 0, 1 and 2 phases,
			with the upper and lower limit
			derived using only securely identified
			type I bursts and including candidates, respectively
			(with a very small difference during mode 2).
			They are plotted against both
			the intra-type II burst average flux $\langle F\sub{pers} \rangle$
			and $\langle F\sub{tot} \rangle$,
			which includes the burst fluence in the average.
			The latter does not match the relation extrapolated from the soft state,
			but it is more compatible with a steeper 
			$t\sub{rec} \propto F\sub{pers}^{-3}$ relation (\textit{dotted-dashed line},
			see Sec.~\ref{sec:trectypeI}.)}}
\label{fig:trec}
\end{figure}

Our data of type I bursts provide a test of the hypothesis
that type II bursts are an accretion phenomenon.
They provide an independent constraint on the accretion rate
by comparing type I recurrence times with
$F\sur{min}\sub{pers}$, $\langle F\sub{pers}\rangle$ and $\langle F\sub{tot}\rangle$.
Fig.~\ref{fig:trec} shows that
the type I burst recurrence time in the hard state
matches the extrapolated fit to the soft state
if $\langle F\sub{pers}\rangle$ is taken
as a proxy of the mass accretion rate,
which traces the persistent emission only.
Adding the burst fluence to yield $\langle F\sub{tot}\rangle$
gives instead too long recurrence times,
even if one includes bursts for which the type I identification
is tentative.
In other words, taking into account the mass that
type II bursts should be accreting on the NS surface
does not yield the predicted recurrence times.
There are three possible explanations for
the mismatch between $t\sub{rec}$ and $\langle F\sub{tot}\rangle$.

Firstly, the type II bursts could be ejection
rather than accretion events.
The disproportionate coincidence in mode 1
of type I bursts with humps during type II burst intermission
would then be explained by having mass accrete onto the NS
only during these intermissions and not during type II bursts.
Ejection is possible in a strong propeller regime,
where the disc is truncated by the magnetosphere at sufficiently
large radii \citep[$r\sub{m} > 2^{1/3}r\sub{cor}$,][]{2014MNRAS.441...86L}.
The mechanism is similar to the one highlighted by \citet[see below]{2010MNRAS.406.1208D, 2012MNRAS.420..416D}
with matter piling up and accreting in burst-like cycles.
In this case, accretion is accompanied
by the launch of a well-collimated,
magnetically dominated Poynting jet.
However, the predicted burst durations and recurrence times
fall in the ms to s range,
and very large $B$ and small mass accretion rates
(well below 10 per cent of the Eddington rate) are necessary.
But the strongest argument against this hypothesis
is that, in a propeller regime,
the luminosity should drastically decrease
with respect to the accreting regime,
because the potential energy that would otherwise be liberated
during the fall onto the NS surface will not be available \citep{2014arXiv1410.3760D}.
As the type II bursts reach luminosities up to the Eddington limit,
we are inclined to reject ejection as a valid hypothesis.

Secondly, the type I bursting rate could respond
not only to the instantaneous mass accretion rate,
but also on the rate at which it itself varies (i.e. $\ddot{m}$).
Possibly, the sudden and short-lived increase in $\dot{m}$
associated with a type II burst
will not produce the same effect in the burning layers
as if the same mass had been more slowly accreted.
To the best of our knowledge, however,
simulations of how the type I burst behaviour
changes with accretion rate only include models
in which $\dot{m}$ is constant over long periods of time
\citep[e.g.,][]{2007ApJ...665.1311H}.
It would perhaps be interesting to see
what these models predict 
in case of swift, large and short-lived changes in the accretion rate,
but we cannot presently elaborate further on this hypothesis.

Thirdly, the shorter duration of type I bursts
in the hard state 
indicates that, as the mass-accretion rate decreases, the RB
approaches a pure-He burning regime.
In this case a steepening of the relation in Fig.~\ref{fig:trec}
is expected. Theoretical models predict
a steeper relation in a pure-He regime \citep{2000ApJ...544..453C},
compatible with the $t\sub{rec} \propto \dot{m}^{-3}$ 
measured in IGR J17480--2446 \citep[see their fig.~11]{2012ApJ...748...82L}.
We plot this empirical relation in Fig.~\ref{fig:trec},
assuming pure He burning takes place for bursts below 10 per cent
of the Eddington limit, at the state transition.
This way we can account for the recurrence times
plotted against $\langle F\sub{tot}\rangle$
during mode 1 and 2, although
type I bursts remain under-abundant for mode 0.

This last hypothesis has the advantage
that it also explains the observed change
in the type I burst duration.
As already mentioned, type I bursts in the hard state
are shorter than those in the soft state,
both in total duration
($\sim 20$~s, versus $\sim 100$~s in the soft state)
and in the duration of their peaks
(see Fig.~\ref{fig:typeIhard}).
This is also ascribable to changes in the fuel composition:
as the mass accretion rate decreases,
the recurrence time of type I bursts lengthens,
leaving less H unburnt by the hot CNO process,
which in turn reduces the burst duration  due to the lack of unstable H burning
via the slow rp-process \citep{1981ApJ...247..267F}.
This means that in the hard state the RB
is approaching a pure-He burst regime.
We think this is the most likely explanation
of the three.

\section{Models for Type II bursts}\label{sec:mod}

Since the discovery of type II bursts in the Rapid Burster, an
assortment of instability mechanisms have been proposed for their
origin. Bursts have been variously attributed to an instability in the
accretion flow (either in viscosity or temperature), instabilities
around the innermost stable circular orbit or as a result of the
interaction between the accretion disc and a stellar magnetic field
(e.g., \citealt{1994A&A...286..491K,1993ApJ...402..593S}). In their
review article, \cite{1993SSRv...62..223L} give an overview of the
various strengths and weaknesses of the different models.
We proceed to review these models again
and test their predictions against 
our observations of the population.

\subsection{Type II bursts: a magnetic phenomenon?}\label{sec:dangelo}

\subsubsection{Theory of disc-magnetosphere interactions}

A weakness of many models proposed to explain type II bursts is the
apparent uniqueness of the RB, since most proposed instabilities
should be widely applicable.  In this respect models based on
magnetospheric instabilities are perhaps the most promising, since a
strong magnetosphere introduces new characteristic lengthscales to the
system, and the distinctiveness of the RB behaviour could then be
attributed to an unusual magnetic field geometry (such as alignment
with the rotation axis, e.g., \citealt{1994A&A...286..491K}), or
unusual relationship between the physical properties that determine
the magnetic regulation of accretion and the properties of the
accretion flow itself (in particular the accretion rate at which the
state transition occurs). On the other hand, since X-ray pulsations
have never been detected in the RB, it is not clear that there is a
strong magnetic field in the system (although by the argument above
this could be due to a near alignment between the rotation and the
magnetic axes).  The BP does have an estimated magnetic field and
measured spin period ($B \sim (2-6) \times 10^{10}{\rm ~G}$, $P
=467$~ms, \citealt{2014ApJ...796L...9D},
\citealt{1996Natur.381..291F}). The observation of type I X-ray bursts
in the RB puts an upper limit of $B \sim 10^{10}{\rm~G}$ for the field
(based on measured fields of know bursters;
\citealt{2012arXiv1206.2727P}), but such a field would still be strong
enough to truncate the accretion disc and regulate the gas flow in the
inner regions of the star at moderate accretion rates (e.g.,
\citealt{1972A&A....21....1P}). Magnetospheric instability models
(studied both analytically and via numerical simulations) have also
received the most theoretical development in the last 20 years (since
the \citealt{1993SSRv...62..223L} review) so we concentrate our
discussion on those results and how they compare with observation.

Accretion on to stars with strong magnetic fields becomes regulated by
the magnetic field once it reaches a certain distance from the star
(the `magnetospheric', or `Alfv\`en' radius, $r_{\rm m}$). At $r_{\rm
  m}$ the disc is truncated by the magnetic field, and the gas is
channelled along magnetic field lines to accrete on the star
\citep{1972A&A....21....1P,1973A&A....24..337S}. The magnetospheric
radius is typically approximated as the point where the ram pressure of the
infalling gas equals the magnetic pressure of the field (with some
considerable uncertainty from the fact that both the disc and the star
are rotating; e.g. \citealt{1979ApJ...232..259G,1993ApJ...402..593S}):
\begin{eqnarray}
\label{eq:rm}
r_{\rm m} &\simeq& 3.1\times10^8 \left(\frac{\dot{M}}{10^{16}{\rm~g~s^{-1}}}\right)^{2/7} \left(\frac{M}{1.4M_{\odot}}\right)^{-1/2}\times\\
\nonumber && \left(\frac{B_*}{10^{10}{\rm~G}}\right)^{4/7}\left(\frac{R_*}{10^6{\rm~cm}}\right)^{12/7}{\rm~cm}.
\end{eqnarray}
The location of the inner disc radius is considerably uncertain and
can vary with the spin rate of the star \citep{1993ApJ...402..593S}. A factor $\sim$$2$
uncertainty for \rin\ should generally be assumed.

Since in general the disc and the magnetic field do not rotate at
the same rate, the coupling between the two will twist the magnetic
field lines, creating a torque that allows for angular momentum
exchange between the disc and the star. The sign of the torque
will depend on the location of \rin\ relative to the {\it co-rotation
  radius}, the radius at which the star's rotation frequency $\Omega_*$
matches the Keplerian frequency of the disc:
\begin{equation}
\label{eq:rc}
r_{\rm c} \equiv \left(\frac{G M_*}{\Omega^2_*}\right)^{1/3}.
\end{equation}

Accretion can only proceed easily when $r_{\rm in} < r_{\rm c}$, that
is, for relatively high accretion rates, slow stellar spin rates, and
low magnetic field strengths. When $r_{\rm in} > r_{\rm c}$ the
stellar magnetic field presents a centrifugal barrier to accretion, and
either expels the infalling gas in an outflow (the `propeller'
scenario; \citealt{1975A&A....39..185I}) or keeps it confined in the
inner disc outside \rc\ (the `trapped disc' scenario;
\citealt{1977SvAL....3..138S,1993ApJ...402..593S,2010MNRAS.406.1208D}).

\subsubsection{Trapped discs}

The trapped-disc picture forms the bases of the most developed model
to explain type II bursts in the RB. In this scenario, when the
accretion rate falls below a certain rate, \rin\ (eq. \ref{eq:rm})
moves outside \rc, and gas is no longer able to accrete on to the
star. As long as \rin\ remains close to \rc, the disc-magnetosphere
interaction will not provide enough energy to unbind the gas from the
disc and launch an outflow. Instead, the gas stays confined in the
disc, gradually piling up as more gas is accreted from larger
radii. This situation continues until enough gas has piled up to push
against the magnetic field for \rin\ to move inside \rc. Here the
centrifugal barrier disappears and the accumulated gas can rapidly
accrete until the reservoir of gas has been depleted and \rin\ again
moves outside \rc. Observationally, the instability will manifest
as short outbursts (as gas accretes freely through the disc and onto
the star) contrasted with periods of non-accretion
(as gas gradually accumulates) with a roughly stable luminosity
generated by the interaction between the disc and the magnetic field.

This instability was described and studied analytically and
numerically in \cite{1993ApJ...402..593S,2010MNRAS.406.1208D}, and
\cite{2012MNRAS.420..416D}, who investigated its general properties
(duty cycle, frequency, outburst shape, etc.) as a function of the
mean accretion rate and the details of the disc-field interaction. In
\cite{2010MNRAS.406.1208D} and \cite{2012MNRAS.420..416D}, the
disc-field interaction is characterized by two (uncertain)
lengthscales: $\Delta r$, the radial extent of the disc that is
coupled to the magnetic field and $\Delta r_2$, the range of \rin\
around \rc\ where the disc changes from being a trapped disc (outside
\rc) and accreting freely (inside \rc). A key feature of this
instability is that it produces bursts on a wide range of timescales,
which can generally be much longer than the dynamical timescale of the
inner disc, and plausibly match the burst timescales seen in the RB.

In the trapped disc instability, the main driver of changes in burst
shape and duration is a changing mean accretion rate into the inner
disc. Roughly speaking, the recurrence time increases and the burst
duration and fluence decrease as the mean accretion rate
decreases. This is broadly seen in the RB: the longest bursts in
mode 0 occur early in the outburst, and the time between mode 2 bursts
increases gradually as the overall luminosity declines.  If this
instability is responsible for the behaviour seen in the RB, the
stochastic behaviour seen in the variation between individual bursts
is mainly attributable to a variable accretion rate in the inner disc
regions.  This may be consistent with the X-ray spectral transition
from a disc-dominated (`soft' or `high') state to a power-law
dominated (`intermediate' to `low') state. In neutron star and
black hole binaries, the spectral transition to a power-law spectrum
is typically accompanied by large luminosity fluctuations on a range
of timescales, which are commonly interpreted as fluctuations in local
mass accretion rate propagated through the accretion flow
\citep[e.g.,][]{1997MNRAS.292..679L}. Since the instability
requires an accumulated mass reservoir, these accretion rate
fluctuations will manifest as a variation in the duration and
frequency of bursts.

However, the simulations of \cite{1993ApJ...402..593S} and
\cite{2010MNRAS.406.1208D} do not show the characteristic `relaxation
oscillator' behaviour seen in type II bursts in the RB, where the wait
time between bursts scales with the fluence of the previous
burst. This is because neither set of simulations properly accounted
for viscosity in the large-scale disc, which sets the timescale to
refill the inner disc regions. Instead the simulations supplied gas
into the inner regions of the disc at a steady rate, which leads to a
steady burst period for a given accretion rate. In a realistic disc,
if the reservoir of gas is emptied before the outer disc can resupply
it with gas, the build-up time will be longer than has been observed
in simulations. This introduces a new timescale into the problem: the
viscous refilling time. If the accretion rate into the inner disc is
unsteady, the timescale between bursts will be variable, but will be
at least as long as it takes for gas to flow into the inner disc. This
may account for the recurrence time-fluence relationship seen in
Fig.~\ref{fig:burst_rels2}: the recurrence time has a minimum (the
viscous inflow rate) but can be much longer (if the supply rate
suddenly decreases). \cite{2012MNRAS.420..416D} indeed found that a
steadily decreasing accretion rate (in a large disc) can induce
chaotic accretion bursts from the trapped disc instability (fig.~10
of that paper). This hypothesis remains to be investigated further.

\cite{2012MNRAS.420..416D} identified two distinct instability regions
for the trapped disk instability: `RI' (long duration, low amplitude
bursts), and `RII' (shorter, smaller amplitude bursts). In `RI'
bursts, $\Delta r_2/r \lesssim 0.02$ (i.e. the disc makes abrupt
transitions between the accreting and non-accreting states), and the
bursts often have complex profiles and with strong contrasts between
bursting and inter-burst luminosity. In contrast, the `RII' bursts
occur at larger values of $\Delta r_2$, bursts are more sinusoidal in
shape with higher luminosities, shorter periods and less contrast
between burst and inter-burst luminosities.  The complex burst
profiles and characteristic outburst timescales of the mode 0 bursts
suggest the RI instability is more likely applicable for the Rapid
Burster.

\begin{figure}
  \resizebox{!}{75mm}{\includegraphics[width=\linewidth]{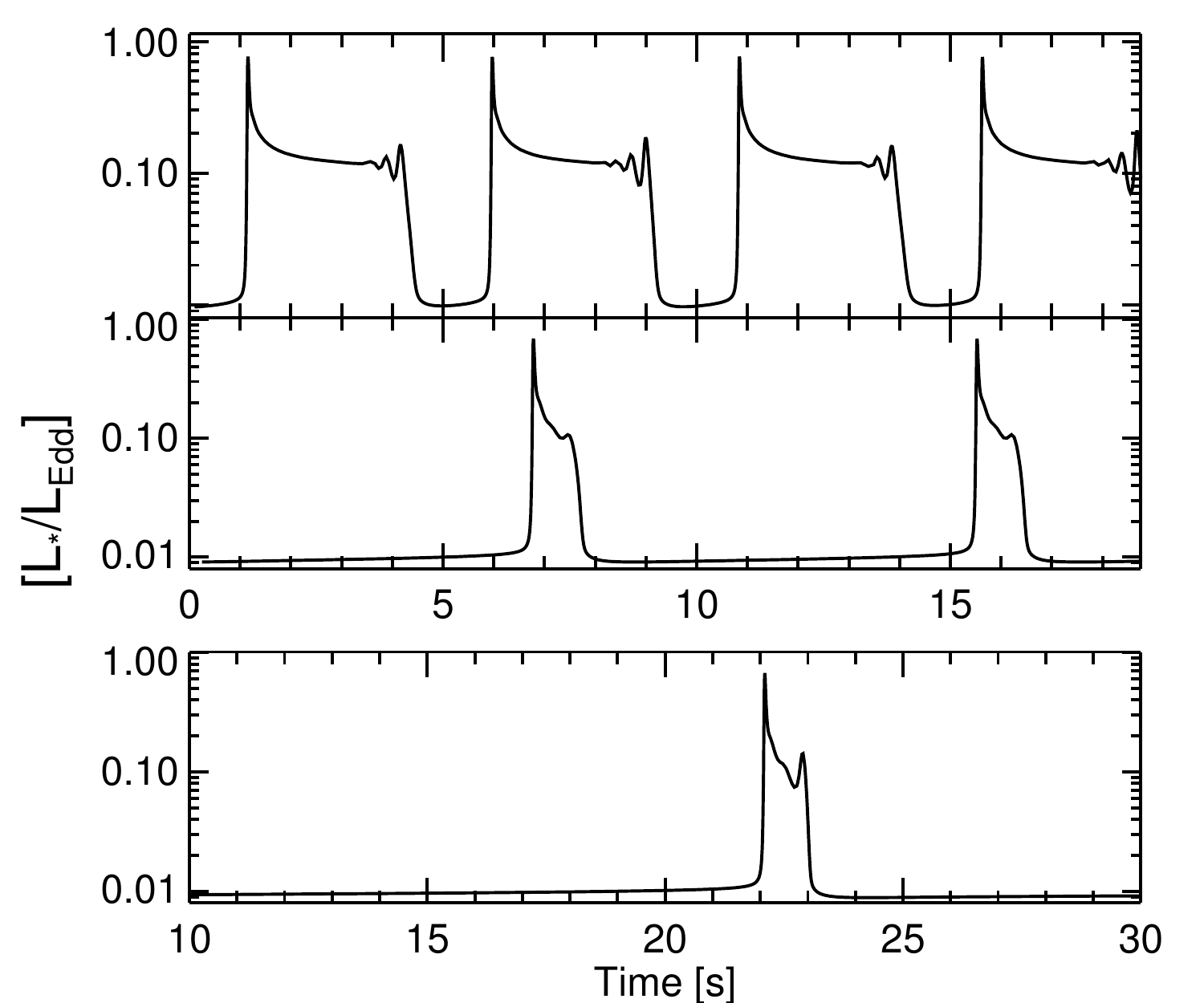}}
  \caption{Bursts predicted by the trapped disc
    instability for a neutron star with $B = 10^{10}$~G and a spin
    period $P = 0.02$~s, for different mean accretion rates (see Sec.~\ref{sec:dangelo}).
    From top to bottom, $\dot{M} = [0.03,0.005,0.002]{\rm ~}\dot{M}_{\rm Edd}$,
    with the burst fluence decreasing ($F =
    [7.8,1.9,1.8]\times10^{38}$~erg) and recurrence time increasing
    as the mean accretion rate decreases.
    We plotted the properties of these simulated bursts
    against those of our sample in Fig.~\ref{fig:burst_rels1}.
    The viscous response of the disc to the changing inner disc may create
    a delay in refilling the inner disc for larger bursts, reproducing
    the behaviour seen in the RB.
    Note that between bursts accretion in the inner disc is suppressed,
    and the disc luminosity is generated by the interaction between the disc
    and the magnetic field.}
\label{fig:bursts_caro}
\end{figure}

At high mean accretion rates, the `RI' accretion bursts are longer,
with burst profiles that frequently show quasi-periodic oscillations
(see Fig.~\ref{fig:bursts_caro}) at the tail of the outburst, similar
to mode 0 bursts of the RB. A flat-topped burst is also sometimes
seen, although typically with an initial spike of accretion, which is
not seen in RB mode 0 bursts. As the mean accretion rate drops, the
burst duration and fluence decreases, and the bursts become more
widely separated in time, similar to evolution in the mode 2 bursts of
the RB (see Fig.~\ref{fig:outburst92026}).

\subsubsection{Application to the RB}

In comparing the predictions of the trapped disc model to the RB
bursts, we consider only the short burst (mode 2 and short mode 1
bursts). This is because the long mode 0 bursts have significantly
different properties: they are often Eddington-limited bursts with a
distinctive burst profile and persistent emission that shows a hump
and (occasionally) type I bursts, suggesting significant accretion
onto the star continues between bursts. We speculate that for mode 0
bursts, the trapped disc instability could appear in conjunction with
a second physical process (perhaps related to the dramatic state
change that marks the onset of type II bursts), but we do not consider
this further.

Applying the trapped disc burst model to the RB is challenging, as its
spin rate and magnetic field are unknown. However, if we assume (as
this model would suggest) that the type II bursts begin when \rin\
moves just outside \rc, then the `critical accretion rate' (that is,
for the onset of bursts) is about 0.1~$\dot{M}_{\rm Edd}$. Equating
\rin\ and \rc\ (equations \ref{eq:rm} and \ref{eq:rc}) at this
accretion rate leads to a degeneracy between $B$ and $P_*$. We can set
some bounds on these parameters from other evidence: the appearance of
type I bursts suggests $B < 10^{10}~$G, while $P_*\gtrsim 0.0017$~s to
make the RB compatible with the fastest-rotating known neutron
stars. This leads to a range of $B\sim 6\times10^7-10^{10}$~G for spin
periods $P\sim2-80$~ms (with the fastest spin periods corresponding to
the smallest magnetic fields). The smaller \rc, the nearer the disc
will extend to the star, and the shorter the recurrence time will be
between bursts, because the burst timescale is generally set by the
viscous diffusion timescale, $t_{\rm visc} \propto r_{\rm in}^2/\nu$
in the disc's inner regions. For an $\alpha$-disc viscosity, $\nu
\sim \alpha(H/r)^2 (GMr)^{1/2}$ (\citealt{1973A&A....24..337S}, where $H$ is
the scale-height of the disc), so that for $\alpha\sim0.1$,
$(H/r)\sim0.1$, $t_{\rm visc} \sim 10$--$1000$~s at \rc\ for the range
of possible spin periods. In
\cite{2012MNRAS.420..416D}, the strongest bursts occur on timescales
0.1--10~$t_{\rm visc}$, which would correspond to 1--100~s if the RB is
a rapid rotator and 100--10$^4$~s if the spin frequency is lower. There
is thus considerable degeneracy in the model, but for a reasonable
range of spin periods, magnetic fields and viscosities, it can
reproduce the observable features of the RB.

In Fig.~\ref{fig:bursts_caro} we show bursts at three different accretion
rates, choosing representative simulated parameters that reproduce the RB range
of luminosity within the constraints on spin and magnetic field set
above. We adopt $B = 10^{10}$~G, $P = 0.02$~s and a viscosity $\nu =
10^{-3} (GMr)^{-1}$, which gives strong bursts when the mean accretion
rate decreases below $\sim10$ per cent $L_{Edd}$. From top to bottom,
the accretion rate decreases: $\dot{M} = [0.03,0.005,0.002]{\rm
  ~}M_{\rm Edd}$. The flux properties of the model agree quite well
with the observed properties of the RB
(see Fig.~\ref{fig:burst_rels1}). The fluence in the bursts
decrease from $7.8\times10^{38}~{\rm erg}$ in the brightest burst to
$1.8\times10^{38}~{\rm erg}$ in the weaker bursts. In comparison, the RB
shows bursts over a somewhat larger range of fluence, with $90$ per
cent of bursts between $0.13$
--$1.8\times10^{39}$~erg. The peak flux in the simulated bursts is
nearly independent of accretion rate, which is also roughly consistent
with the RB, where the peak luminosity of 90 per cent of the bursts
spans a range of $\sim2$. 

As discussed in Sec.~\ref{sec:persistent}, the persistent
emission between bursts does not show obvious correlations with other
burst properties, and varies by a factor $\sim2$ for the majority of
bursts. This is naturally accounted for by the trapped disc model,
where between bursts accretion in the inner disc is suppressed, and
the disc luminosity is generated by the interaction between the disc
and the magnetic field, which adds angular momentum and energy to the
disc. Since the location of the inner disc does not vary much between
bursts, the luminosity also stays roughly constant, regardless of the
fluence of the subsequent burst. As mentioned above, this does not
account for the `humps' and occasional type I bursts seen between long
mode 0 bursts, since the distinctive nature of mode 0 bursts suggests
additional physics beyond the theoretical picture outlined above.

The observed range of burst recurrence times and durations are less
well accounted for in the model. As described above, all simulations
of the instability assumed a constant accretion rate in to a small
region of the inner disc, which does not properly account for the time
it could take the disc to refill after an outburst. As a result, the
recurrence time scales linearly with accretion rate -- the lower the
mean accretion rate the longer it takes to refill the reservoir and
the weaker the subsequent burst. This is the opposite of what is
apparently observed in the RB. However, if the accretion rate in the
inner disc is strongly variable (as is postulated to explain rapid
luminosity variability in the hard state;
\citealt{1997MNRAS.292..679L}), the relationship between the mean
accretion rate and burst fluence is less well-defined. Simulations
also show bursts with a smaller variation in burst durations (about a
factor of 4, with durations between $\sim (2-8)$~s) than the RB,
in which the majority of bursts (excluding mode 0, which as we argue
above likely have an additional physical process governing their
behaviour) last between 2--24~s, with a peak around 3~s.

In summary, the trapped disc instability can reasonably account for
multiple properties of the RB short type II bursts for a variable
accretion rate in the inner disc regions. Simulations can match
observed burst fluences, maximum luminosities, and persistent flux
properties for reasonable (albeit poorly constrained) spin and
magnetic field. The range of simulated burst durations and fluences
are somewhat smaller (by a factor $\sim4$ and $\sim2$ respectively)
than is observed in the RB. Most significantly, the relaxation
oscillator behaviour is not reproduced, although as we argue above,
this is at least in part because the response of the large-scale
viscous disc to the burst instability has not been properly
modelled. The finite viscosity in the disc will delay the refilling
time once gas is depleted, lengthening the time between bright bursts.
The RB uniqueness may be explained by a special combination of
magnetic field strength and stellar spin period (which together sets
the accretion rate where the instability can manifest) and an
alignment between the magnetic field and the spin axis.

\subsection{Other magnetospheric instabilities}
Several other mechanisms have been proposed to produce instabilities
in the accretion flow that could explain the RB behaviour, but it is
not clear that the timescales on which they are expected to operate
will match the RB. Both mechanisms outlined below rely on the ideal MHD
properties of the accretion flow, i.e. that field lines are frozen
in to the plasma and may not efficiently re-couple to the disc once
the connection is severed (which is expected to happen rapidly due to
differential rotation between the field and the disc). This can lead
to either compression of the magnetic field by the disc moving inwards
or unstable fingers of accretion onto the neutron star surface via
interchange instabilities. 

\cite{1994A&A...286..491K} proposed that the perfectly conducting disc
can be completely shielded from the magnetic field of the star, so
that the disc will spiral inwards and pinch the dipolar magnetic field
along the equator into a distorted `butterfly' shape. This compression
lasts until the matter touches the surface of the star, whereupon the
gas penetrates through the field and accretes on to the star. This
releases the pressure on the magnetic field, which rapidly re-expands
and restarts the cycle. Some version of this scenario is seen in all
magnetic accretion simulations
(e.g. \citealt{1997ApJ...489..890M,1997ApJ...489..199G,1996ApJ...468L..37H,2009MNRAS.399.1802R}),
which use ideal MHD and hence shield the disc from the magnetic
field. However, due to the finite numerical diffusivity present in all
simulations, the timescale for the instability is typically a few
dynamical timescales, which is much shorter than the RB bursts ($\sim
50-1000$ dynamical times, depending on the spin of the
star). Moreover, it is not at all clear theoretically that gas will
have difficulty coupling to the magnetic field, as is required by this
picture. Simulations of MRI turbulent discs with stratified density
profiles (e.g. \citealt{2010ApJ...713...52D}) show the disc field
extends vertically out of the disc, which could form a low-density
magnetized corona above the disc surface
(e.g. \citealt{2014ApJ...784..169J}). This would provide an obvious
site for reconnection, which could recouple the disc to the star and
prevent strong compression of the magnetic field.

In this scenario, the burst should have a gradual rise (as the
accretion disc spirals inwards toward the star) with a sudden spike
when the gas finally hits the surface. This also seems inconsistent at
least with the short bursts, which show low persistent emission
between bursts, although it could perhaps explain the inter-burst humps
seen between mode 0 bursts. In order to match the observed
bursting timescales, the original \cite{1994A&A...286..491K} paper
also posited a strong $10^{12}$~G field for the RB, which is
inconsistent with the observation of type I X-ray bursts at relatively
high luminosities.

An alternative scenario will develop when the field is not strong
enough to efficiently disrupt the disc, allowing accretion to proceed
via interchange (Rayleigh-Taylor) instabilities on to the star. In
this picture the RB magnetic field could be significantly weaker than
assumed above (or the spin rate lower), so that the disc extends down
to the star until the accretion rate drops to $\sim10$ per cent
$L_{\rm Edd}$. At lower luminosities the field will truncate the disc
but not enough to enforce accretion on to the poles, and fingers of
accretion will extend down to the stellar surface. This was studied in
depth in the in the 3-D simulations of Kulkarni \& Romanova
\citep{2008MNRAS.386..673K}. However, as in the above scenario, the
timescale for the instability is typically the orbital timescale of
the inner disc edge, which will be smaller than the spin period of the
star and thus likely considerably shorter than the observed burst
timescales unless the RB is a very slow rotator.

\subsection{Thermal/viscous instability models}
A number of models have also been proposed that
do not rely on a role for the magnetic field.
\citet{1993SSRv...62..223L} have already reviewed them
and explained their shortcomings;
we will briefly discuss them in the light
of our sample properties.

A class of models invokes thermal \citep{1976MNRAS.175..613S}
or viscous instabilities \citep{1974ApJ...187L...1L}
in the accretion disc to explain the luminosity fluctuations.
Both instabilities arise in the inner portion of an $\alpha$ disc,
where the radiation pressure dominates
and electron scattering is the main source of opacity.
Thermal instabilities are due to an inability to maintain
a thermal balance, while viscous instabilities
arise because the viscous stress becomes inversely proportional
to the surface density, breaking up the inner disc in concentric rings
as a result.

These analytical results however only established \textit{local}
criteria for instabilities to grow.
They were first tested in a global, time-dependent
simulation by \citet{1984ApJ...287..761T},
who confirmed the existence of unstable solutions.
The produced bursts liberate
about the right amount of energy (\erg{36-38}),
and the observed durations (0.1--1~s)
and recurrence times ($\sim$ 10~s)
at least overlap with the (however much broader)
ranges observed in type II bursts.
Also, the instability develops in a range of mass accretion rates
($(0.1-0.45) \times 10^{18} \textrm{~g~s}^{-1}$,
between 10 and 45 per cent $L\sub{Edd}$),
above which the flow becomes stable again.

There are two main problems with this model.
Firstly, different time scales and energetics
can only be reproduced with variations of the global
mass accretion rate.
Although this problem is shared by the class of magnetospheric models
discussed earlier, we should stress that at least in that case
a much broader variety of burst properties is observed,
closer to matching the ranges of fluences, durations and recurrence times
that we have characterized in this study.
Secondly, models relying on disc instabilities leave
the uniqueness of the RB unaddressed,
as they should apply to all accreting compact
objects, even black hole binaries,
and have indeed been put forward to explain the variability of
sources like Cyg X-1.

\subsection{GR models}

An entirely different approach was taken by
\citet{1992ApJ...385..651W}, who proposed that
the RB be an exceptional source in hosting a NS with a radius
smaller than the innermost stable circular orbit (ISCO),
surrounded by a massive accretion disc.
The latter gives the very low viscosity the model requires.
Also, the magnetic field must be weak enough
not to disrupt the accretion flow.
In this case, rather than having the magnetosphere act as a gate,
radiation torques inside the ISCO lead to a sudden spike in the accretion rate,
that repeats after the inner disc has been replenished on a viscous time scale.

The predicted burst luminosities match those observed in type II bursts,
and the bursts last between roughly 1 and 50~s.
A very large disc would further be compatible with the large orbital period
of the BP ($P\sub{orb} = 11.8$~d), while no orbit information
is available for the RB.
The model suffers from the same problem that afflicts all other models,
namely the lack of a relaxation oscillator behaviour.
But its main shortcoming is that it cannot account
for type II bursts in the BP, which was only discovered
three years after this model was proposed.
\citet{1992ApJ...385..651W} claim that the
presence of a massive disc means that even relatively large magnetic fields
(up to $10^{12}$~G) would not disrupt the Keplerian disc before it crosses
the ISCO.
Although estimates of the magnetic field strength in the BP
are as low as  $(2-6)\times 10^{10}$~G \citep{2014ApJ...796L...9D},
there cannot be any doubt that magnetospheric accretion is happening in that source,
because of the presence of pulsations at the spin frequency,
both in the persistent emission and in the bursts.
If GR effects were truly responsible for type II bursts in the RB,
then another mechanism altogether would be at play in the BP.

\section{Conclusions}
Our population study has highlighted
a number of new, important features of the type II bursts.
The bursts are Eddington limited,
which leaves little doubt as to their origin lying
in accretion episodes.
We have constrained the duration of these
instabilities to cover the $10^{-1}-10^3$~s range,
and we have shown that a minimum recurrence time
of order of 10~s seems necessary to develop a burst.
We encourage dropping the old mode 0, 1 and 2 classification,
adopting instead a simpler one between short and long bursts,
which reflects the dichotomy these show
in their interaction with the persistent emission.

We judge models based on thermal and viscous instabilities
too general to explain the scarcity of sources
displaying type II bursts, while models based
on GR effects imply that a different phenomenon altogether
is responsible for the BP bursts.
Models based on the gating role of the magnetosphere
can reproduce a number of properties in our sample,
while at the same time requiring a fine tuning between
the stellar spin, the magnetic field strength, the mass accretion
rate and the alignment between the magnetic field and the spin axis,
and thus account better for the uniqueness of the RB.

\footnotesize{
  \bibliographystyle{mn2e}
  \bibliography{paper}
}

\clearpage

\appendix

\section{Burst search algorithm}
\label{sec:appendix_search}

The thousands of type II bursts from the RB in the RXTE archive
necessitate an automatic burst search algorithm for the identification
process. We now proceed to explain the algorithm that we developed
from trial and error. During the development, results were verified by
visual inspection of the light curves.

The automatic procedure employs a light curve generated at 1~s
resolution from all PCA '\textsc{Standard-1}' data with the RB in the FOV (2.4~Msec),
for all photon energies and for PCU2 which is the
only PCU that is always on. Furthermore, the pointing and the
elevation of the source above the Earth's horizon was tracked. From the
pointing, the off-axis angle between the optical axis to the RB and 4U
1728-34 were calculated. Pointing and elevation information is
available at 16-s resolution. For unsampled times, the pointing
information was taken from the prior data point.

First, for each light-curve bin an estimate was made of the
persistent emission underlying a potential burst at that bin,
by determining the average of all
flux values within 100~s prior to that bin, provided that the pointing
is constant within 0.05~deg for at least the previous 5~s.
If there was no reliable background from those data, the average
flux was taken from the posterior 20 to 100~s interval. The background
thus determined was subtracted from the flux of each bin.
Second, a
'significance' was determined by taking the ratio to the Poisson error
of that bin. If the significance was larger than 1, the bin was further
investigated for the presence of a burst.
Third, the
background-subtracted flux of subsequent bins after that bin were added
to the signal and the accrued significance determined. This 
process was repeated
for as long as the significance increased, allowing for it to decrease only
once. When finished, an attempt was made to accrue signal going
backward in time from the initial bin.
Fourth, the candidate burst was
verified by checking that no other bursts previously determined (i.e.,
starting at earlier times) were in the new burst interval, that the
total significance was at least 10, that the elevation was at least 2.5
degrees, that the pointing was constant during the burst and that there
was at least one 'empty' bin between two consecutive bursts. A list of
candidate bursts was thus composed. Since this allows for a more careful
background determination from the original flux array by excluding the
burst intervals and, fifth and last, the procedure was repeated.
When it was impossible to determine the background, the burst was ignored.

All type I bursts from the RB show
the typical fast-rise-exponential-decay (FRED) shape.
Soft-state type I bursts have rise times (7.1~s)
and durations ($\sim$100~s; B13),
which are relatively long for type I bursts,
and typical of the burning of H-rich fuel \citep{1993SSRv...62..223L}.
As we have shown in Sec~\ref{sec:typeIhard},
hard-state type I bursts are typically shorter ($\sim$40~s).
Nonetheless, type II bursts of comparable durations
cannot mimic the FRED shape,
because multiple peaks appear in their decays \citep{1991MNRAS.251....1T}.
This allows identification of a RB type I burst even when
its profile in the light curve
overlaps with those of several fast-recurring type II bursts
(as in Fig.\ref{fig:tI_1vs2}, bottom).

As for type I bursts from 4U 1728-34,
they show instead the much faster time scales
(rise time $<$1~s, duration $\simeq$10~s)
that are typical of H-poor nuclear fuel \citep{2010ApJ...724..417G},
and based only on this could be confused with short RB type II bursts.
However, the bursts from 4U 1728-34
show a characteristic bimodal distribution of peak fluxes
($9.2 \times$\ergscm{-8} for the bursts showing photospheric radius expansion (PRE)
and $4.5 \times$\ergscm{-8} for normal bursts, \citealt{2008ApJS..179..360G}),
which are much larger than observed in RB type II bursts.
Still, when the PCA is pointed at the RB the peaks fluxes
of the weakest 4U 1728-34 bursts can be similar to those
of the brightest RB type II bursts.
Again, the multiple peaks in the decays reveal the nature of the latter.

Thus, a list of 8754 bursts was derived. Eighteen of these concern
overlap between two kinds of bursts. The breakdown of three kinds of
bursts is: type I bursts from the RB (123, 9 overlapping), type-I
bursts from 4U 1728-34 (167, 9 overlapping) and type II bursts from
the RB (8458, 18 overlapping). Twenty-two bursts could not be
definitely identified because of insufficient statistics. They are
either type I from the Rapid Burster or type II. Otherwise, type-I
bursts from 4U 1728-34 are easily recognizable because they are
approximately two times brighter than the brightest type II bursts
from the RB and last the shortest. The latter characteristic is
necessary for identification because, when the PCA points at the RB
and 4U 1728-34 is an off-axis angle of 0.5~degrees, the collimator
shadow cuts the measured peak flux in roughly two and is comparable to
the brightest bursts of the RB.  For the other two pointing
configurations, identifying bursts from 4U 1728-34 is unambiguous. The
difference between type I and II bursts from the RB is also
manageable. Type I bursts are at least 20~s long, have a typical long
tail
\citep{2013MNRAS.431.1947B,2014MNRAS.437.2790B}
and show no ringing, contrary to most type II
bursts.

The start and end times of the bursts were determined by searching,
starting from the bin with the highest flux, downward and upward,
respectively, for the last times when the flux drops below 10\% of the
peak flux. Notably, we identified a record duration for a mode-0
burst. Observation ID 92026-01-06-03 apparently starts off during a
type II bursts which subsides only after 1100~s in a fashion typical
for a mode-0 burst.

\section{Spectral analysis}
\label{sec:spectralstudy}

In principle one would like to perform a full time-resolved
spectroscopic analysis of each type II burst to determine parameters
such as peak flux (in erg~cm$^{-2}$s$^{-1}$), burst fluence (in erg~cm$^{-2}$), fluence
of the persistent emission between bursts and $\alpha$. However, given
previously published spectral analyses of RB data
data \citep[e.g.,][]{1979ApJ...227..555M,1995MNRAS.277..523R,2001AJ....122...21M,2003PASJ...55..827M},
there is merit in asking
whether that is necessary. Spectral variability during and between
type II bursts is presumably limited and the average energy per photon
is probably nearly constant. In order to verify this and determine a
conversion factor between PCA-detected photons and energy, we
analyzed a representative subset of the data. The subset was drawn
from the 'offset' observations that are free from contamination by 4U
1728-34. Eleven observations, identified by unique ObsIDs, were
chosen representing typical as well as extreme behavior of the
RB. They are listed in Table~\ref{tabspeclist}.

\begin{table*}
\begin{center}
\caption[]{A representative subset of RB observations for spectral
  studies\label{tabspeclist}}
\begin{tabular}{lclc}
\hline \hline
ObsID & Date of & Type of behavior & Bolometric flux range\\
      & Observation & &  (10$^{-8}$~erg~cm$^{-2}$s$^{-1}$)\\
\hline
20418-01-01-01R & 17-jun-1997 & bright soft state & 1.2 \\
20093-01-07-01R & 26-jun-1997 & high soft state & 1.3 \\
20418-01-07-00  & 10-jul-1997 & fainter soft state & 0.26 \\
30424-01-01-02R & 25-aug-1998 & mode 0, short bursts & 0.1-1.7 \\
30424-01-02-02R & 29-aug-1998 & mode 1, sequence of 5 bursts & 0.1-2.5 \\
30424-01-03-02R &  1-sep-1998 & mode 1, sequence of 20 bursts & 0.1-2.8 \\
30424-01-04-02R &  4-sep-1998 & mode 2, sequence of 22 bursts & 0.1-0.8 \\
30424-01-06-01R & 10-sep-1998 & mode-2, single burst in tail outburst & 0.1-0.8 \\
92026-01-06-00  & 23-jun-2006 & mode 0, sequence of 5 short bursts (brightest fluxes from RB) & 0.2-3.3 \\
92026-01-06-03  & 26-jun-2006 & mode 0, record-long burst ($>$1300 s) & 0.15-0.8 \\
92026-01-06-04  & 27-jun-2006 & temporary return to soft state & 0.45-0.55 \\
\hline\hline
\end{tabular}
\end{center}
\end{table*}

The time resolution of the spectral analysis varied across the
different observations and within observations is chosen to follow the
variability in flux. In total 374 spectra were extracted from these 11
observations, involving 8 mode-0 bursts (including the longest one
detected), 25 mode-1 bursts and 23 mode-2 bursts.

The spectral analysis involves the following generalities. Spectra
were extracted from event-mode data with 64 spectral channels. All
active PCUs were employed. The bandpass was limited to 2.5-20 keV
which is where the PCA is well calibrated
\citep{2006ApJS..163..401J,2012ApJ...757..159S} and contains most of the signal. 
A systematic error of 0.5 per cent per channel was
included \citep[following][]{2012ApJ...757..159S}. All spectra were corrected for
particle background and cosmic diffuse background as predicted through
the {\sc pcabackest} tool (version 3.8). The PCA response matrix was
calculated with version 11.7.1 of tool {\sc pcarsp}. This includes a
correction for the collimator response which is always non-standard
because the RB is always considerably off-axis in the selected
ObsIDs. The collimator throughput for the RB was usually about
0.4. Spectral bins were combined to obtain at least 20 counts per bin
to ensure applicability of the $\chi^2$ statistic.  All spectral
modelling was carried out with {\sc xspec} version 12.8.2. A constant
cold interstellar absorption component was assumed to apply,
equivalent to $N_{\rm H}=1.6\times10^{22}$~cm$^{-2}$
\citep{1995AJ....109.1154F}.
The absorption model of \citet{1983ApJ...270..119M} was followed.

\begin{figure}
\includegraphics[width=0.98\columnwidth,angle=0]{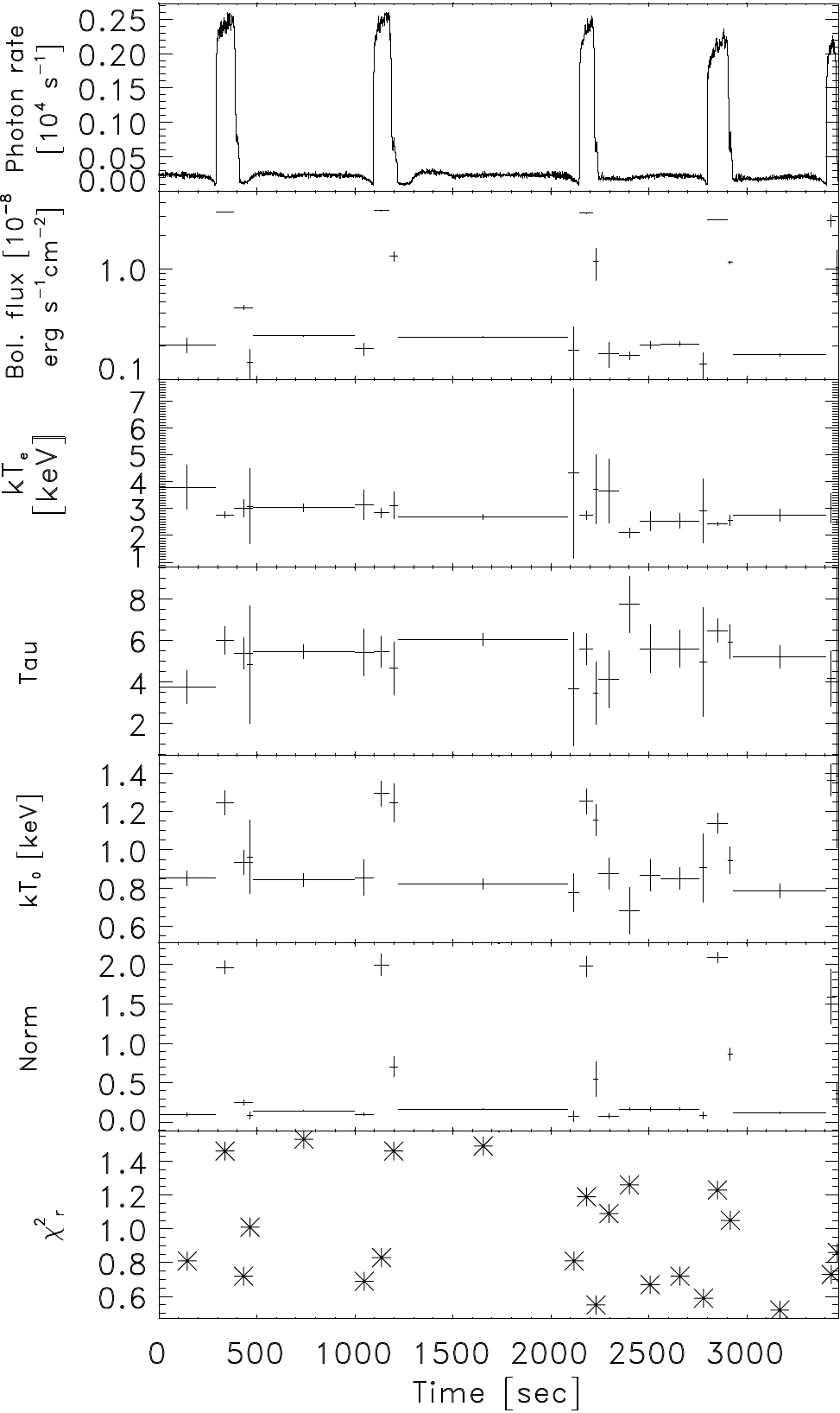}
\caption{Time-resolved spectroscopy for ObsID 92026-01-06-00. The response
is corrected for the off-axis position of the RB. The fitted model is 
the Comptonization model according to \citet{1994ApJ...434..570T}.
\label{spec_exampleK}}
\end{figure}

A significant contribution may be expected from the Galactic Ridge. To
determine this, off-axis observation ID 92026-01-04-00 was chosen
which represents the observation with the lowest flux measured for the
RB field.  It was taken on June 9, 2006, 11 days before the onset of
the outburst presented in Fig.~1. The background-subtracted data was
modelled with the combination of a Raymond-Smith component \citep{1977ApJS...35..419R}
and a power law, following the prescription of \citet{1998ApJ...505..134V} as also
determined from PCA ridge data. The fit was satisfactory with
$\chi^2_\nu=1.79$ ($\nu=20$). This constant spectral component was
assumed to be present in all studied RB spectra.

We tested various models on the data: single component models (power
law, Planck function, thermal bremsstrahlung, Comptonization) and
models with combinations of two components. It was found that models
with three free parameters were sufficient to describe all spectra. We
chose the Comptonization model according to 
\citet{1994ApJ...434..570T,1995ApJ...450..876T}
as baseline since this is the simplest model fitting all
data. Furthermore, we find that this model performs best to the
highest quality spectrum (20093-01-07-01R whose spectrum contains
10$^6$ counts), with $\chi^2_\nu=0.92$ for $\nu=20$. As an example,
for the same data $\chi^2_\nu=11.3$ for a model consisting of a disk
black body and a power law, and 6.88 for a cut-off power law.  We
note that it was necessary to complement the continuum model with a
narrow line at about 6.5 keV, in the spectral region where the Fe-K
resonance line resides (6.4-6.9 keV, depending on the ionization state
of Fe).
Possibly it is related to instrumentally
scattered emission from the bright source 4U 1728-34 just outside the
field of view,
or to an inadequate modelling of the Galactic Ridge at this location.

Figure~\ref{spec_exampleK} shows an example of a time history for
observation 92026-01-06-00 which shows a sequence of 5 bright and
relatively short mode-0 bursts. What is clearly visible here, and that
applies in general to all our spectral results, is that plasma
temperature and optical depth show no significant variability, while
photon seed temperature does. The plasma temperature varies in general
between 2.5 and 3.5 keV while at the same time the plasma optical
depth varies between 7 and 4, both typical 1-sigma errors of 1 in
value.
A more detailed analysis of the spectral data is subject for a future paper.
We here confine the analysis to inferring an average translation factor
between photon fluxes and fluences and energy fluxes and fluences.

For the most accurately determined 112 spectral shapes, we find that a
photon flux of 1~c~s$^{-1}$PCU$^{-1}$ for the whole bandpass
translates on average to a 2.5-20 keV energy flux of
$(1.02\pm0.02)\times10^{-11}$~erg~cm$^{-2}$s$^{-1}$. All 112 conversion values are
contained within 7\% from this average. Thus, the maximum systematic
error by assuming a constant conversion factor is 7\%. Converting from
2.5-20 keV energy flux to unabsorbed bolometric flux involves a factor
of $1.24\pm0.07$. All 112 bolometric correction factors range between
-7\% and +2\% from the mean. Therefore, the maximum systematic error
in the bolometric correction factor is 7\%.

\begin{figure}
\includegraphics[width=\columnwidth]{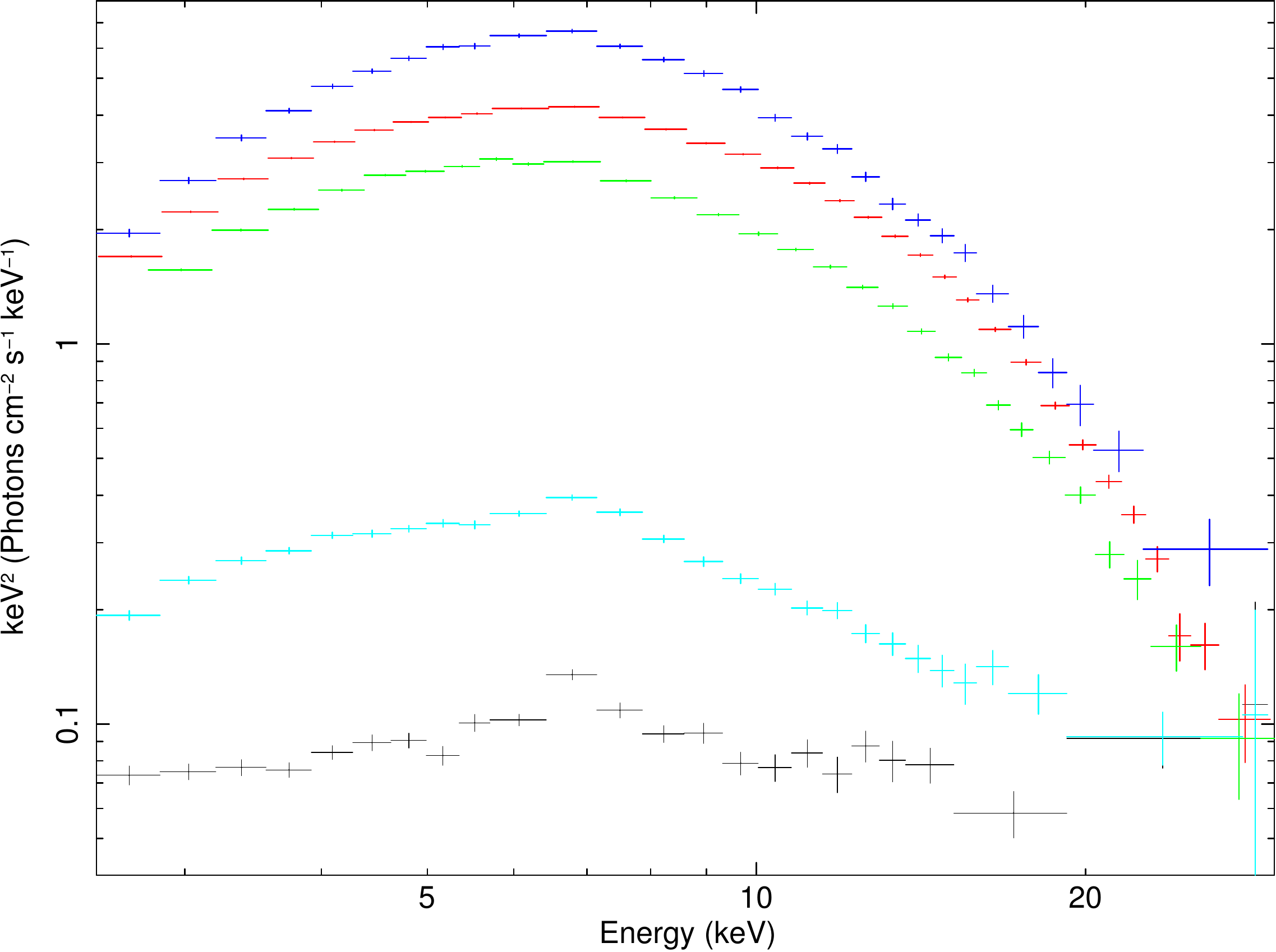}
\caption[]{Unfolded 2.5-30 keV $\nu-f_\nu$ spectra for five widely
  different states of RB. From bottom to top: non-burst emission
  between slow mode-2 bursts (black curve, ObsID 30424-01-06-01R),
  non-burst emission in long intermission between mode-1 bursts (light
  blue, 30424-01-02-02R), peak of longest mode-0 burst ever measured
  (green; 92026-01-06-03), soft state prior to mode-0 (red;
  20418-01-01-01R), peak of bright short mode-0 burst (dark blue;
  30424-01-01-02R). The lowest spectrum shows the iron line clearly
  whose origin is undetermined.
\label{spec_spectra}}
\end{figure}

In Fig.~\ref{spec_spectra} we show spectra of the five most diverse
states of the RB. The conversion factor from c~s$^{-1}$PCU$^{-1}$ to
unabsorbed bolometric erg~cm$^{-2}$s$^{-1}$\ varies between $1.23\times10^{-11}$ and
$1.56\times10^{-11}$.

\label{lastpage}
\end{document}